\documentclass[11pt]{article}
\usepackage{jheppub}
\usepackage{amsmath,amssymb,amsfonts,graphicx,slashed,amsthm,mathtools,upgreek, enumerate, tensor, subfig}
\usepackage[dvipsnames]{xcolor}
\usepackage{arydshln}

\usepackage{comment}
\usepackage{hyperref}
\usepackage[utf8]{inputenc}
\usepackage[titletoc]{appendix}

\numberwithin{equation}{section} 
\allowdisplaybreaks

\allowdisplaybreaks

\newcommand{\be}{\begin{equation}}
\newcommand{\ee}{\end{equation}}
\newcommand{\f}{\frac}
\newcommand{\s}{\sqrt}
\newcommand{\p}{\partial}

\newcommand{\bea}{\begin{eqnarray}}
\newcommand{\eea}{\end{eqnarray}}
\newcommand{\ba}{\begin{align}}
\newcommand{\ea}{\end{align}}

\newcommand{\ve}{\varepsilon}

\newcommand{\la}{\langle}
\newcommand{\ra}{\rangle}
\newcommand{\beq}{\begin{equation}}
\newcommand{\eeq}{\end{equation}}

\title{
Evaporation of black holes in flat space entangled with an auxiliary universe
}

\author[a]{\! Akihiro Miyata,}
\author[b,c]{\! Tomonori Ugajin}

\affiliation[\,a]{Institute of Physics, University of Tokyo, Komaba, \\ Meguro-ku, Tokyo 153-8902, Japan}
\affiliation[\,b]{Center for Gravitational Physics,
Yukawa Institute for Theoretical Physics, Kyoto University,
Kitashirakawa Oiwakecho, Sakyo-ku,
Kyoto 606-8502, Japan}
\affiliation[\,c]{The Hakubi Center for Advanced Research, Kyoto University,
Yoshida Ushinomiyacho, Sakyo-ku, Kyoto 606-8501, Japan}
\emailAdd{miyata@hep1.c.u-tokyo.ac.jp}
\emailAdd{tomonori.ugajin@yukawa.kyoto-u.ac.jp}

\preprint{UT-Komaba/21-3,\;
YITP-21-27}

\abstract{
We study a  thermofield double type entangled state on two disjoint universes $A$ and $B$,  where  one of the universes  is asymptotically  flat  containing a black hole.  
As we increase  the  entanglement temperature, this black hole receives back-reaction from the stress energy tensor of the state.   This results in lengthening of the wormhole region in the black hole interior, and decreasing of its  horizon area, both of which are key features of an evaporating black hole. 
We then compute the entanglement entropy  on the universe $A$ through the island formula, and argue that it naturally follows the Page curve of an evaporating black hole in flat space.  We also study the effects of local operations in the gravitating universe with the black hole.  We find that they accelerate the evaporation of the black hole, therefore disrupt the entanglement between  two universes. Furthermore, we observe that depending on whether the  operation can be regarded as  an LOCC or not, the behavior of the entanglement entropy changes.   In particular, when the operation is made neither in the entanglement wedge of the radiation system or that of the black hole, the transition between  the island phase and the no-island phase  can happen multiple times.
}


\makeatletter
\gdef\@fpheader{}
\makeatother
\begin{document}

\maketitle

\parskip=10pt

\section{Introduction}

Recently  it has been claimed that in the presence of semi-classical gravity,  the island formula gives a correct  prescription to compute entanglement  entropy \cite{Almheiri:2019hni, Almheiri:2019psf,Penington:2019npb,Penington:2019kki,Almheiri:2019qdq}.  This formula is inspired by the holographic entanglement entropy formula \cite{Ryu:2006bv,Ryu:2006ef,Hubeny:2007xt}  and its quantum corrections \cite{Faulkner:2013ana, Engelhardt:2014gca} in the AdS/CFT correspondence. When this formula is applied to a semi-classical black hole which is evaporating due to Hawking radiation, the entropy of the radiation naturally follows the Page curve \cite{Page:1993wv,Page:2013dx}. This  provides a resolution of   the black hole information loss problem, and  implies that semi-classical gravity is consistent with  unitarity of quantum theory.  The way this new rule provides the correct entropy involves a region called ``island" in the black hole.  This island region naturally arises, when we compute the entanglement entropy using a gravitational path integral through the replica trick \cite{Penington:2019kki,Almheiri:2019qdq}.  In fact,  the rule to evaluate a  gravitational path in the semi-classical limit appears to include all saddles consistent with given boundary conditions. It was argued that there is an overlooked gravitational saddle  in the Hawking's calculation of the radiation entropy. This new saddle is called a replica wormhole, and this saddle gives the dominant contribution after the Page time. 
See for a review of this topic \cite{Almheiri:2020cfm}. 

One way to efficiently study black hole evaporation is, introducing another  auxiliary universe, say universe $A$ which we assume to be non-gravitating, and consider an entangled state  $|\Psi \ra_{AB}$ on $A$ and the original  gravitating universe $B$ with the black hole\footnote{Another way to efficiently study  the evaporation process  is to holographically realize  the system,  by introducing branes on which gravitational degrees of freedom are living
\cite{Almheiri:2019hni,Penington:2019kki, Balasubramanian:2020hfs,Rozali:2019day, Chen:2020uac, Chen:2020hmv,Akal:2020twv, Geng:2020fxl,Kawabata:2021hac,Geng:2021wcq,Fallows:2021sge,Anderson:2021vof,Li:2020ceg, Akal:2020ujg,Deng:2020ent, Bhattacharya:2021jrn}.}.  One can think of this new system being generated out of  the  system in the single universe with an evaporating black hole,  by gathering all  Hawking quanta, and sending them to the auxiliary universe. Therefore the entanglement of  the $|\Psi \ra_{AB}$  mimics the one of the Hartle-Hawking state on the evaporating black hole.   This setup has   been used to  study the entropy of Hawking radiation  of  two dimensional black holes in anti-de Sitter (AdS) space and de Sitter  in JT gravity \cite{Balasubramanian:2020coy,Balasubramanian:2020xqf}. See for studies with a similar approach \cite{Penington:2019kki,Hartman:2020khs}. Other applications of the island formula to de Sitter space can be  found in \cite{Chen:2020tes,Hartman:2020khs,Sybesma:2020fxg,Geng:2021wcq}  Indeed the entanglement entropy of this new system on AB naturally follows a  Page curve as a function of the entanglement temperature. 

In this paper, we generalize this analysis to black holes in  flat space.  Evaporation of such  two dimensional  asymptotically flat black holes, as well as time evolution of their  radiation entropy have been studied since early 90's \cite{Callan:1992rs,Fiola:1994ir,Russo:1992ht}. Previous  applications of  the island formula to black holes in asymptotically flat space can be found, for example in \cite{Hashimoto:2020cas,Hartman:2020swn, Anegawa:2020ezn, Krishnan:2020oun, Gautason:2020tmk, Matsuo:2020ypv, Wang:2021woy, Wang:2021mqq}. Since the total state  $| \Psi \ra_{AB}$ induces  the stress energy tensor expectation value $\la \Psi |T_{ab}| \Psi \ra$ to the gravitating universe, the black hole in this universe receives  back-reaction from it.  As in the cases of dS and AdS black holes in JT gravity, we show that this  back-reaction  is crucial to get the Page  curve  for the  black hole in flat space.  Two key effects of the    back-reaction on the black hole are the following. First, it makes the wormhole in the interior region longer, second  it reduces the horizon area. This implies, when we compute the  entanglement entropy  $S(\rho_{A})$ of the non-gravitating universe using the island formula, it starts to decrease   when the  entanglement temperature  is increased above some  threshold value. This is in contrast to the case of AdS black holes, where  the  entropy is  saturating to some constant value.  The decrease of the entropy  is plausible, because a black hole in flat space   evaporates through Hawking radiation, and  loses its entropy.

We believe our setup clarifies several ambiguities in the previous discussions on the island formula applied to  asymptotically flat black holes.  In the previous discussions, 
the radiation subsystem (the heat bath) is often naively taken to be the region $R$ located far away from the black hole, but it is still gravitating. However, such a choice for the radiation region $R$ is worrisome because of several reasons.   First, there is no consistent way to defined a ``region" in the presence of gravity, in a diffeomorphism invariant manner. In addition, the Hilbert space of quantum gravity never has a factorized into the Hilbert space on $R$ and its complement, due to the edge modes on the boundary of the region $R$. Second, when the heat bath is gravitating, the naive island formula is no longer valid, as has been shown in the recent papers \cite{Geng:2020fxl,Geng:2021iyq,Anderson:2021vof, Balasubramanian:2021wgd}. This is because in the setup there is a novel wormhole connecting the gravitating bath and the black hole appears in the path integral for the R\'enyi entropies. This is a concrete realization of  the ER =EPR \cite{Maldacena:2013xja} slogan, which predicts the existence of such a wormhole connecting the early radiation and the black hole interior.

These two concerns are avoided in a very clear manner in our setup of two disjoint universes. Namely, in our setup since the radiation region is located on the non-gravitating universe A which differs from the gravitating universe with the black hole, there is no ambiguity in defining the radiation region from the first place. This in particular means that our definition of the entanglement entropy is unambiguous. Also in our setup, since two universes are disjoint  we can safely turn off gravity in one of the universes, thus we do not need to worry about the existence of the wormhole connecting the bath and the black hole.

We then addressed the question  of whether the entropy computed from the island formula is operationally meaningful from  the quantum information theoretic point of view.To this end we study perturbations of the black hole  by local operations in the gravitating universe. One way to model such a local operation is applying  an operator 
well localized in the region of the interest, and this protocol  is known as a local quench.   Time evolution of the entanglement entropy  in such quench processes have been extensively studied, for example in \cite{Calabrese:2007mtj,Caputa:2014vaa,Nozaki:2013wia,Asplund:2014coa,Caputa:2015waa, Ugajin:2013xxa,Asplund:2013zba,Nozaki:2014hna,Nozaki:2014uaa,Caputa:2014eta,David:2016pzn}.  We  find
such local operations ``accelerate" the evaporation, especially when the operators are inserted in the interior of the black hole. Effects of such shock waves on the radiation entropy 
has been studied in  \cite{Hollowood:2020cou, Almheiri:2019psf, Goto:2020wnk,Chen:2019uhq, Chen:2020jvn,Hollowood:2020kvk} using the island formula.

We indeed find that our results obtained  from the island formula are  operationally meaningful. For instance 
when the insertion is made  either in the entanglement wedge of the radiation or in the similar wedge of the black hole, then the resulting entanglement  entropies  always decreases. This is consistent with the interpretation that such a local operator insertion is an LOCC (local operations and classical communications.)

This paper is organized as follows.  In section \ref{section:setup}  we  explain our setup, and the previous results on the island formula in this setup in  detail. In section \ref{section:flatBH}, we  introduce the black hole solution of our interest, and study how it is deformed due to  the back-reaction of the stress energy tensor expectation value of the total state. We then use the island formula to compute the entanglement entropy of  the non-gravitating universe A. The result naturally follows a Page curve of an evaporating black hole. In section \ref{section:BHshock}, we study the Page curve in the presence of local operations in the gravitating universe.   After classifying possible quantum extremal surfaces, we discuss  effects of the local operations to the entanglement entropy.   We conclude this paper in section \ref{section:conc}. 
Appendix \ref{section:EEderivation} we review and discuss  time dependence of  the entanglement entropy in a local quench in a two dimensional conformal field theory with a large central charge.

 \section{Setup}
 \label{section:setup}
 
 \subsection{Two disjoint asymptotically flat  universes}
 
 Let us first explain the setup we consider in this paper. First of all,   we prepare two disjoint universes, say $A$ and $B$ which are asymptotically flat (see figure \ref{fig:SetupMin}). We then  define two identical CFTs, on each universe  $A$ and $B$. For simplicity, in this paper, we only consider two  dimensional spacetimes.  Furthermore, we turn on semi-classical gravity on the universe $B$.  Thus the effective action of each universe  reads
 \be
 \log Z_{A} =\log Z_{{\rm CFT}},\quad \log Z_{B} =-I_{{\rm grav}} +\log Z_{{\rm CFT}}.
 \ee
 
 As the gravitational part $I_{{\rm grav}}$ of the above effective action, we choose the  ${\rm CGHS}$ action \cite{Callan:1992rs}, 
 \be 
 I_{{\rm grav}}=    \f{1}{4\pi}\int dx^{2} \s{-g}  \left(\Phi R -\Lambda \right), 
 \label{eq:action}
\ee
which is a theory of gravity in two dimensional asymptotically flat space.
 This action involves two fields, namely dilaton $\Phi$ and metric 
$g_{\mu \nu}$.  Also, we introduced   an auxiliary parameter  $\Lambda$, for the later purpose\footnote{The reader should not confuse this parameter $\Lambda$ with a cosmological constant.}.
 
 The total Hilbert space of this system  is naturally bipartite $H_{A} \otimes H_{B}$.  Since these two universes $A$ and $B$ are disjoint, they can not exchange classical information, but states on $A$ and $B$ can be entangled quantum mechanically. In this paper, we are interested in the structure of the entanglement of the states on the bipartite Hilbert space.  To study this concretely, 
 on the bipartite system we will consider the thermofield double (TFD) state 
 \be 
 |\Psi \ra = \sum_{i=0}^{\infty} \s{p_{i}}\; | i \ra_{A} \;  | \psi_{i} \ra_{B},  \quad p_{i} = \f{e^{-\beta E_{i}}}{Z(\beta)}, \label{eq:TFDstate}
 \ee
 where $Z(\beta)$ is a normalization factor which makes sure the condition $\la \Psi | \Psi \ra =1$,  $| i \ra_{A} $ is an energy eigenstate of the CFT on the non-gravitating universe A, and $| \psi_{i} \ra_{B} $ is the same energy eigenstate on the gravitating universe $B$.  Although they are identical states, since  gravity is acting on the universe B, and  properties of the state is affected by this,  so  we instead write them  differently.  $\beta$ in \eqref{eq:TFDstate} characterizes the amount of the entanglement in this state. For this reason, $1/\beta$ is sometimes called the entanglement temperature.

 \subsection{Islands in the setup}

In our previous papers \cite{Balasubramanian:2020coy,Balasubramanian:2020xqf}, we studied the entanglement entropy $S(\rho_{A})$  of the TFD state \eqref{eq:TFDstate} on the non-gravitating  universe $A$.  This quantity is defined by the von  Neumann entropy
\be
S(\rho_{A}) =-{\rm tr} \rho_{A} \log \rho_{A}, \quad \rho_{A} = {\rm tr}_{B} | \Psi \ra \la \Psi |  \quad.
\ee

This von  Neumann entropy is computed by the replica trick, ie first compute the R\'enyi entropy ${\rm tr} \rho_{A}^{n}$,  which has a path integral representation  on the gravitating universe $B$, then at the end of the calculation send $n \rightarrow 1$.  This gravitational  path integral  is evaluated in the semi-classical limit $G_{N} \rightarrow0$, by including all saddles consistent with given boundary conditions. In particular,  by taking into account a wormhole which connects all replicas (replica wormhole), we obtained the following formula for the entanglement entropy,
\be
S(\rho_{A}) = {\rm min} \{S_{{\rm no-island}},  S_{{\rm island}}\}.
\label{eq:defent}
\ee

$S_{\rm no-island}$ in the above formula coincides with  the CFT thermal entropy $S_{\rm no-island} =S_{\beta} (B)$,  
\be
S_{{\rm th}} (B) = -\sum_{i} \;p_{i} \log  p_{i},
\ee
with $p_{i}$ defined in\;\eqref{eq:TFDstate}.   The other contribution $S_{{\rm island}}$ in the above formula is given by taking the extremum of the generalized entropy,
\be
S_{{\rm island}} =\underset{\bar{C}}{\rm Ext}   \left[\Phi (\p \bar{C})  + S_{\beta}[\bar{C}] -S_{{\rm vac}}[\bar{C}] \right] \equiv \underset{\bar{C}}{\rm Ext}  \; S_{{\rm gen}}[\bar{C}] 
\label{eq:generalizedentropy}
\ee
over all possible  intervals  $\bar{C}$ in the
gravitating universe $B$.  $\Phi (\p \bar{C}) $ is the ``area term" of the generalized entropy, which is in the current case given by the sum of the  dilaton values at the boundary of the interval $\bar{C}$.   Also, $S_{\beta}[\bar{C}]$  is the entanglement entropy of  thermal states on $\bar{C}$, and $S_{{\rm vac}}[\bar{C}]$ is the  entropy of the vacuum state.     Since the TFD state is pure on the total system AB,  the generalized entropy  satisfies  $S_{{\rm gen}}[\bar{C}]   = S_{{\rm gen}}[AC]$.   This implies  the interval $C$ in the gravitating universe $B$ can be identified with the island in our setup.

We are  interested in the behavior of the  entanglement entropy $S(\rho_{A})$ as we tune the entanglement temperature $1/\beta$, especially when the gravitating universe $B$ contains a black hole. It was argued that \cite{Balasubramanian:2020coy,Balasubramanian:2020xqf},  in the low temperature regime   $\beta \gg 1$, since $S_{{\rm no-island}}< S_{{\rm island}}$ this entanglement entropy \eqref{eq:defent} coincides with the thermal entropy  $S_{{\rm th}}(B)$, which is an analogue of  the Hawking's result for the radiation entropy of  evaporating black holes. Also, this implies the entropy is linearly growing, as we increase the entanglement temperature $1/\beta$. At sufficiently high temperature $S_{{\rm no-island}}$ is larger than $S_{{\rm island}} $. According to the formula \eqref{eq:defent}, in this regime the entanglement entropy is given by $S_{{\rm island}}$, instead of the naive Hawking's entropy $S_{{\rm no-island}}$. Furthermore, in this limit,
$S_{{\rm island}}$ almost coincides with  the entropy of the black hole in the gravitating universe B. This is how the Page curve of an evaporating black hole is reproduced in the current setup.

 \subsection{Embedding of two universes}

\begin{figure}[t]
    \centering
    \includegraphics[width=6cm]{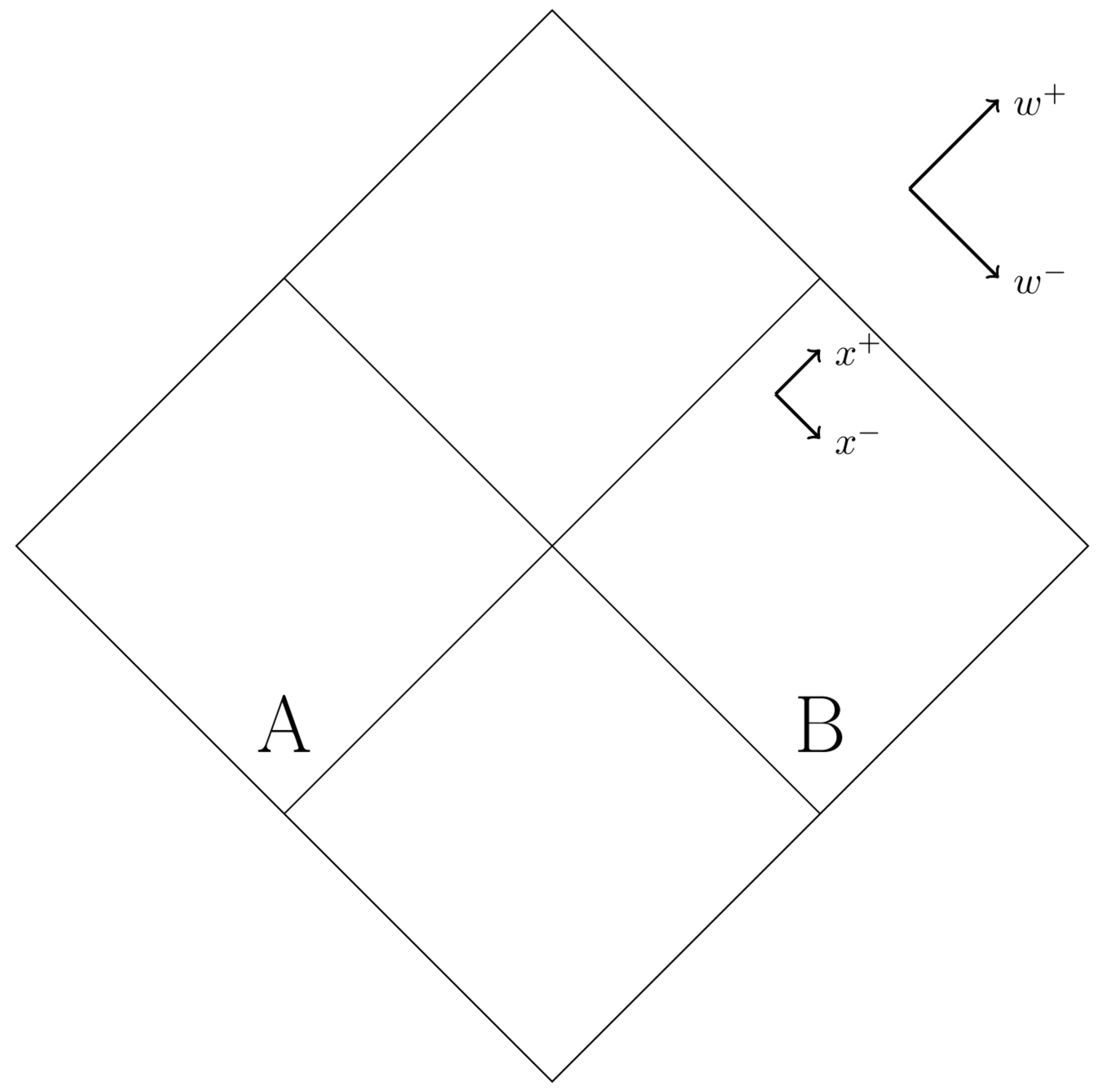}
    \hspace{0.5cm}
  \hspace{0.5cm}
    \caption{\small{ We consider a system with two disjoint asymptotically Minkowski spaces, A and B. In this figure, these universes are embedded in a larger Minkowki space.   }}
    \label{fig:SetupMin}
\end{figure}

 One way to study the setup is to embed the system to a larger Minkowski space $M$, as in figure \ref{fig:SetupMin}. 
Each universe is a wedge in the larger space. The non-gravitating universe is the left wedge of $M$ and similarly the gravitating universe is the  right wedge.  To be more specific,  let us define the light-cone coordinates $x^{\pm} =x \pm t$ on each universe. Also, let $(w^{+},w^{-})$ be coordinates of the larger Minkowski space  $M$. The embedding map is defined by 
\be 
w^{\pm} =e^{\f{2\pi}{\beta}\; x^{\pm}}. 
\ee

The non-gravitating universe $A$ is mapped to the left wedge of $M$, $w^{\pm} <0$  and the gravitating universe $B$ is mapped to the right wedge of $M$, $w^{\pm}>0$.   Also the thermofield double state on AB is mapped to the global Minkowski vacuum of $M$.

 \section{An asymptotically flat black hole and  its radiation entropy}
 \label{section:flatBH}

 The purpose of this paper is to study a similar entropy in asymptotically flat spacetime, using the island formula \eqref{eq:defent}. To do so, we need to specify the dilaton  profile $\Phi$  which appears in the generalized entropy \eqref{eq:generalizedentropy}.  
 Since  the thermofield double state induces the thermal  stress energy tensor expectation value  $\la \Psi |T_{\pm\pm} |\Psi \ra$ on the gravitating universe $B$, this dilaton receives back-reaction from it.

In the ${\rm CGHS}$ model with the action \eqref{eq:action},  the metric is always fixed to the flat one, as the variation of the action with respect to $\Phi$ sets $R=0$. 
We will find it convenient to use  the compact coordinates $(x^{+}, x^{-})$ , in which the flat  metric is given by 
\be 
ds^{2} = -\f{dx^{+} dx^{-}}{\cos^{2} x^{+}\cos^{2} x^{-}}, \quad -\f{\pi}{2} \leq x^{\pm} \leq \f{\pi}{2}.
\ee
These  coordinates are related to the standard coordinates $(X^{+},X^{-})$
 with the metric $ds^{2} = -dX^{+} dX^{-}$, by  $X^{\pm} = \tan x^{\pm}$.
In the coordinate system,
$x^{+} =\pm \f{\pi}{2}$
 and $x^{-} =\pm \f{\pi}{2}$ correspond to asymptotic infinities  of the spacetime.

The equations of motion for the dilaton is given by 
\be 
\nabla_{a} \nabla_{b}\Phi -g_{ab}\nabla^{2} \Phi = \f{\Lambda}{2} g_{ab} -2\pi  \; \la \Psi | T_{ab} | \Psi \ra.
\ee

In general, in the  light-cone gauge where the metric takes the form,
\be
ds^{2} =-e^{2\omega} dx^{+} dx^{-} ,
\ee
these equations of motion are reduced to 
\be 
-e^{2\omega} \p_{\pm} \left[ e^{-2\omega} \p_{\pm} \right] \Phi= 2\pi\; \la \Psi | \;T_{\pm \pm} | \Psi \ra, \quad 
\p_{+} \p_{-} \Phi=  2 \pi \;\la \Psi | T_{+-} | \Psi \ra +\f{\Lambda}{4} e^{2\omega}  .  \label{eq: EOMs}
\ee

\begin{figure}[t]
    \centering
    \includegraphics[width=5cm,scale=1]{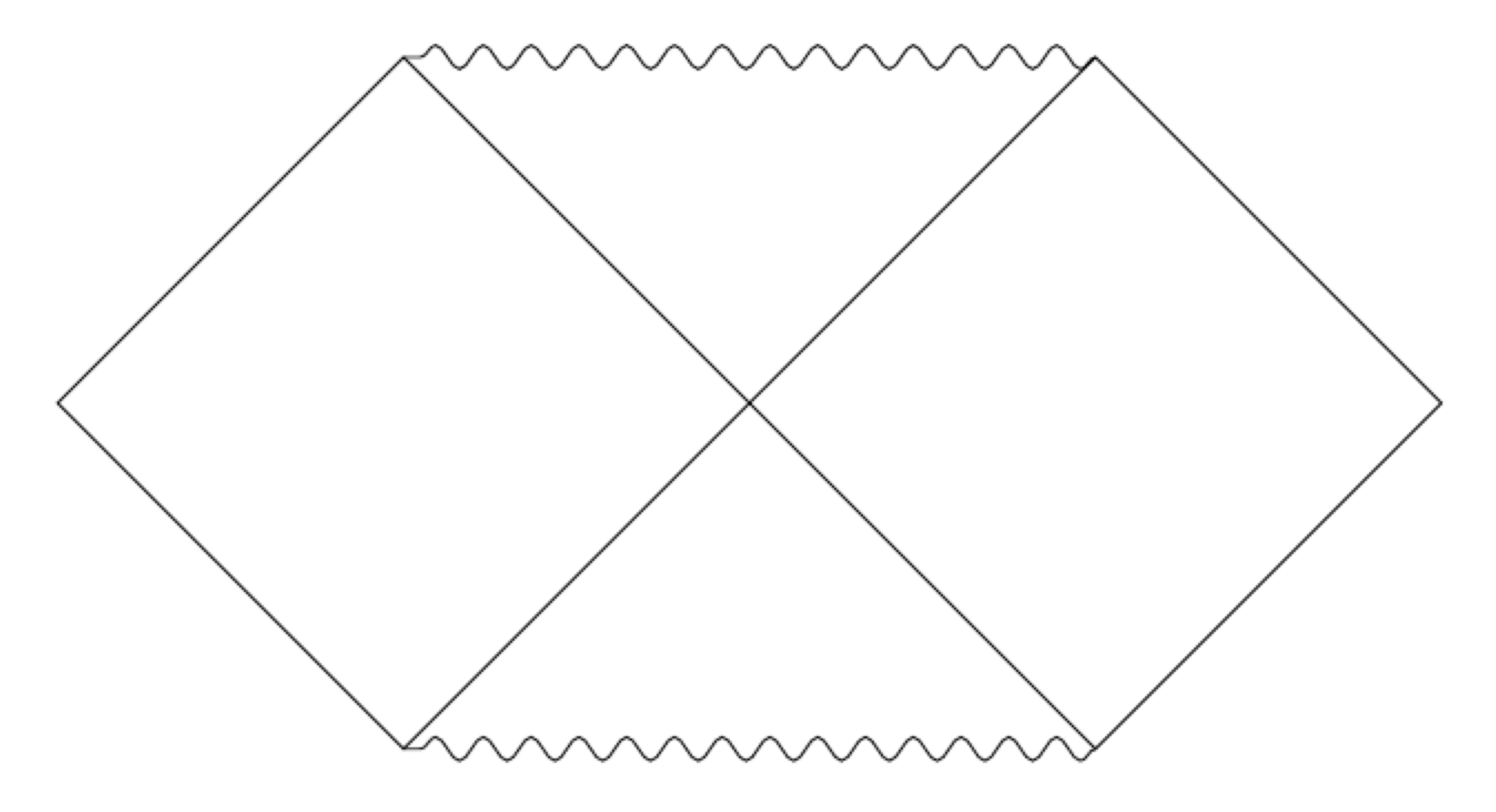}
    \hspace{0.5cm}
\includegraphics[width=8cm]{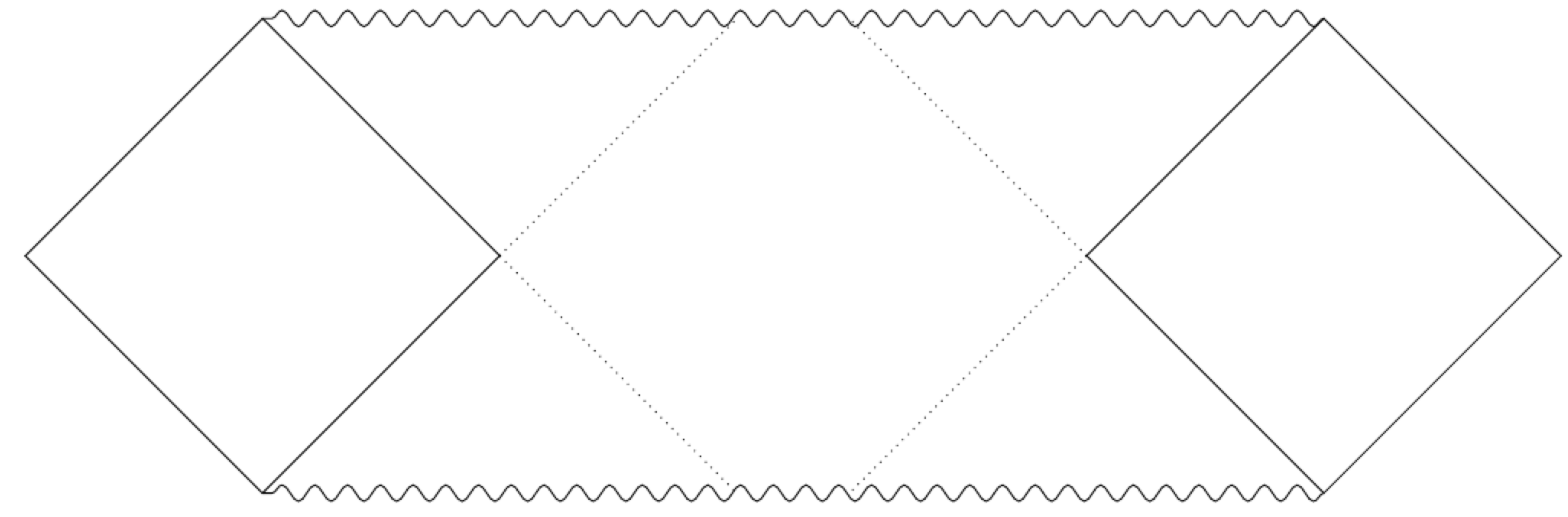}
  \hspace{0.5cm}
    \caption{\small{{\bf Left}: The Penrose diagram of the black hole without the back-reaction. {\bf Right}: The Penrose diagram of the black hole with the back-reaction of the source
  (\ref{eq:stress}). It develops a long wormhole region in its interior. }}
    \label{fig:blackholes}
\end{figure}

\subsection*{The Sourceless solution}

Let us first discuss the dilaton profile  when  the stress energy tensor  is vanishing $ \la \Psi |T_{ab}| \Psi \ra =0$.  In our setup, this happens when the  entanglement temperature is low, $\beta \rightarrow \infty$.  It reads
\be 
\Phi_{0} = \phi_{0} + \f{\Lambda}{4} \tan x^{+} \tan x^{-}, 
\label{eq:simpledil}
\ee
where $\Lambda$ is the parameter in the ${\rm CGHS}$ action  \eqref{eq:action}. 
As we will see, this dilaton profile corresponds to an asymptotically flat eternal black hole, whose Penrose diagram is identical to the standard 4d Schwarzchild eternal black hole
(refer to the left panel of Fig. \ref{fig:blackholes}).

\subsection*{The Solution with the source}

As we increase the entanglement temperature, it is no longer possible to  neglect the back-reaction of the stress energy tensor to the dilaton profile. The stress energy tensor expectation value  of the thermofield double state \eqref{eq:TFDstate} is,
\be
\la \Psi | T_{\pm \pm}| \Psi \ra= \f{c}{24\pi} \left( \f{2\pi}{\beta}\right)^{2}.
\label{eq:stress}
\ee

By solving the equations \eqref{eq: EOMs} for $\Phi$, we get 
\be 
\Phi_{\beta}=\phi_{0} + \f{\Lambda }{4}\tan x^{+} \tan x^{-} -X_{\beta} \left(x^{+} \tan x^{+} +x^{-} \tan x^{-}\right), \quad X_{\beta}\equiv \f{c}{24} \left( \f{2\pi}{\beta}\right)^{2}\;  . 
\label{eq:dilbeta}
\ee

This solution corresponds to an eternal black hole with a  long interior region (the right panel of Fig. \ref{fig:blackholes}).

\subsection{Penrose diagrams}

Now having presented the dilaton profile of our interest \eqref{eq:dilbeta}, let us discuss the causal structure of the spacetime described by the profile.   Since it turns out that it  corresponds to an eternal black hole, we are interested in the location of the singularity and the event horizon.  A useful fact is that,  the dilaton is vanishing $\Phi=0$ at the black hole singularity. Also,  the  bifurcation surface of the black hole  is a critical point   $\p_{\pm} \Phi=0$. The entropy of the black hole is given by the dilaton value at the critical point.

\subsection*{The sourceless solution}

As a warm-up, let us describe the causal structure of the dilaton profile \eqref{eq:simpledil} without the source. 
In this case, the location of the singularity satisfies 
\be 
\Phi_{0}=0 \leftrightarrow \tan x^{+} \tan x^{-} =-\f{4\phi_{0}}{\Lambda}. 
\ee
In the standard coordinates $(X^{+}, X^{-})$,  this singularity is just a hyperbola $X^{+} X^{-} =-4\phi_{0}/\Lambda$, which is expected.
This singularity intersects with the right future null infinity $x^{+}=\f{\pi}{2}$ at $x^{-}=0$. Similarly, it intersects with the left future infinity $x^{-} =-\f{\pi}{2}$ at $x^{+} =0$.  This fixes the location of the event horizon  to $x^{+}=0$ and $x^{-} =0$. Indeed, 
this black hole has only one 
 bifurcation surface, ie, at  $x^{\pm} =0$.  The value of the dilaton at the bifurcation surface is
$\Phi(0) =\phi_{0}$, 
which is equal to  the entropy of the black hole, and is independent of $\Lambda$.

\subsection*{The solution with the  source} 

The structure of the spacetime   is eventually deformed  by turning on the stress energy tensor \eqref{eq:stress},   due to the back-reaction, which is described by the dilaton profile $\Phi_{\beta}$ \eqref{eq:dilbeta}.
We can read off  the location of the   singularity  in the deformed spacetime  from $\Phi_{\beta}$. Near the 
 the right future null infinity $x^{+} =\f{\pi}{2}$,  the dilaton profile  is  approximated as
 \be
 \Phi_{\beta} = \f{\Lambda}{4} \tan x^{+} \left(\tan x^{-}  -\f{2\pi X_{\beta}}{\Lambda}\right), \quad x^{+} \rightarrow \f{\pi}{2}. 
 \ee
 Therefore, the singularity intersects with the future infinity at $x^{-} =x^{-}_{c}$ with,
\be 
\tan x^{-}_{c} =\f{2\pi X_{\beta}}{ \Lambda}.
\label{eq:defxcm}
\ee

As we increase the entanglement temperature $\beta \rightarrow 0$,  $X_{\beta}$ in the right hand side gets large, and the intersection  approaches spatial infinity,  $x^{-}_{c} \rightarrow \f{\pi}{2}$ with $x^{+} =\f{\pi}{2}$.   Similarly, the singularity intersects with the left future null infinity  $x^{-}=-\f{\pi}{2}$ at $x^{+} =x^{+}_{c}$  with 
\be 
\tan x^{+}_{c} =-\f{2\pi X_{\beta}}{\Lambda},
\label{eq:defxcp}
\ee
again in the high temperature limit, it satisfies  $x^{+}_{c} \rightarrow -\f{\pi}{2}$, so this intersection approaches the  opposite spatial infinity.  Since the dilaton profile \eqref{eq:dilbeta} is  invariant under time reflection $x^{+}  \leftrightarrow x^{-}$,  the singularity intersects with the past null infinity in a similar fashion.  Namely it intersects with the right past null infinity $x^{-} =\f{\pi}{2}$ at $x^{+} =-x^{+}_{c}$ with \eqref{eq:defxcp} and the left 
past null infinity $x^{+} = -\f{\pi}{2}$
at  $x^{-} =-x^{-}_{c} $ with \eqref{eq:defxcm}.   In summary, as one increases the entanglement temperature,  the singularity of the black hole comes closed to the reflection  symmetric slice  $x^{+} = x^{-}$.

This also fixes the location of the event horizon of the black hole. The right future horizon of the  black hole is at  $x^{-} =x^{-}_{c}$ with \eqref{eq:defxcm}. Similarly the left future horizon is at $x^{+} = -x^{+}_{c} $. Since these two future horizons do not intersect  on the reflection symmetric slice $x^{+} =x^{-}$, this  black hole contains a region in its interior, which is causally inaccessible from asymptotic infinities (the right panel of figure \ref{fig:blackholes}) . Such a region is called a causal shadow region.  The fact that the black hole singularity approaches the  reflection symmetric slice as we increase the entanglement temperature means the causal shadow region gets larger and larger in this limit.

One can also confirm this by finding  locations of the bifurcation surfaces $(x^{+}_{H}, x^{-}_{H})$ which satisfy $\p_{\pm} \Phi_{\beta}=0$. Because of the  symmetry $x^{+} \leftrightarrow x^{-}$ of the dilaton profile \eqref{eq:dilbeta}, these bifurcation surfaces satisfy $x^{+}_{H}=x^{-}_{H} \equiv y$, and 
\be 
\f{\Lambda}{4}\tan y -X_{\beta} \left( \cos y \sin y +y \right) =0. \label{eq:bifdilbeta}
\ee

We are interested in the $\beta \rightarrow 0$ limit, where the 
 two solutions $y =y_{\pm}$ of the equation  the equation \eqref{eq:bifdilbeta}  satisfy
\be 
\tan y_{\pm} = \f{4X_{\beta} y_{\pm}}{\Lambda}.\label{eq:ces woShock}
\ee

Both of these bifurcation surfaces get close to the spatial infinity, 
$y_{\pm} \rightarrow \pm \f{\pi}{2}$. 
The dilaton value  at these bifurcation surfaces in the same  limit is given by 
\be
\Phi_{\beta} (x^{\pm}_{H})=\phi_{0}-\frac{(\pi X_{\beta})^{2}}{\Lambda}.\label{eq:dilatonvCES}
\ee

Notice that the dilaton value  $\Phi_{\beta} (x^{\pm}_{H})$ at the horizon  decreases as we increase $1/\beta$.
This means that, as we increase the  entanglement between the two universes $\beta \rightarrow 0$,  the back-reaction  of the CFT stress energy tensor makes the horizon area of the  black hole  smaller.  This  is closely related to  fact that quantum mechanically,   a black hole  in flat space   evaporates  by emitting of Hawking  quanta. Indeed  our setup can be regarded as  an idealization of  the black hole plus a radiation system .  The radiation degrees of freedom is modeled by the CFT degrees of freedom in our setup, and the entanglement of the CFT thermofield double state  between $A$ and $B$  is the avator of the entanglement in the Hartle-Hawking state.  Therefore the  increase of  the entanglment  of the TFD state  (which we do  by hand ), captures the late time physics of   the actual black hole evaporation process, and as a result, the black hole in our setup loses its entropy.

We can not have a semi-classical description of the black hole, at the very final stage of the evaporation. This is because, as we increase the entanglement temperature, both 
future and past singularities  come close to the reflection symmetric slice, and eventually touch the slice.  This critical temperature can also be read off  from the dilaton values at the bifurcation surfaces \eqref{eq:dilatonvCES}, where it becomes zero.

\subsection{Quantum extremal surface}

\begin{figure}[t]
    \centering
    \includegraphics[width=7cm]{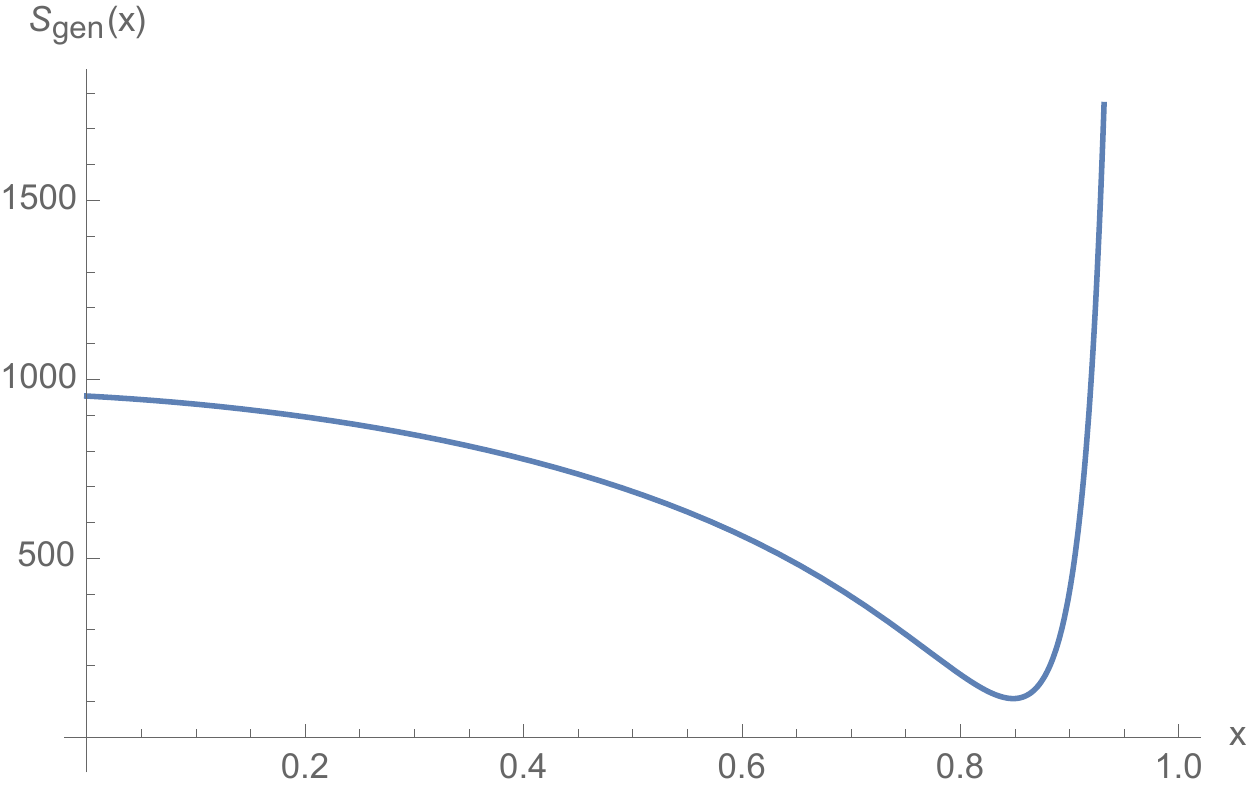}
    \hspace{0.5cm}
\includegraphics[width=6.5cm]{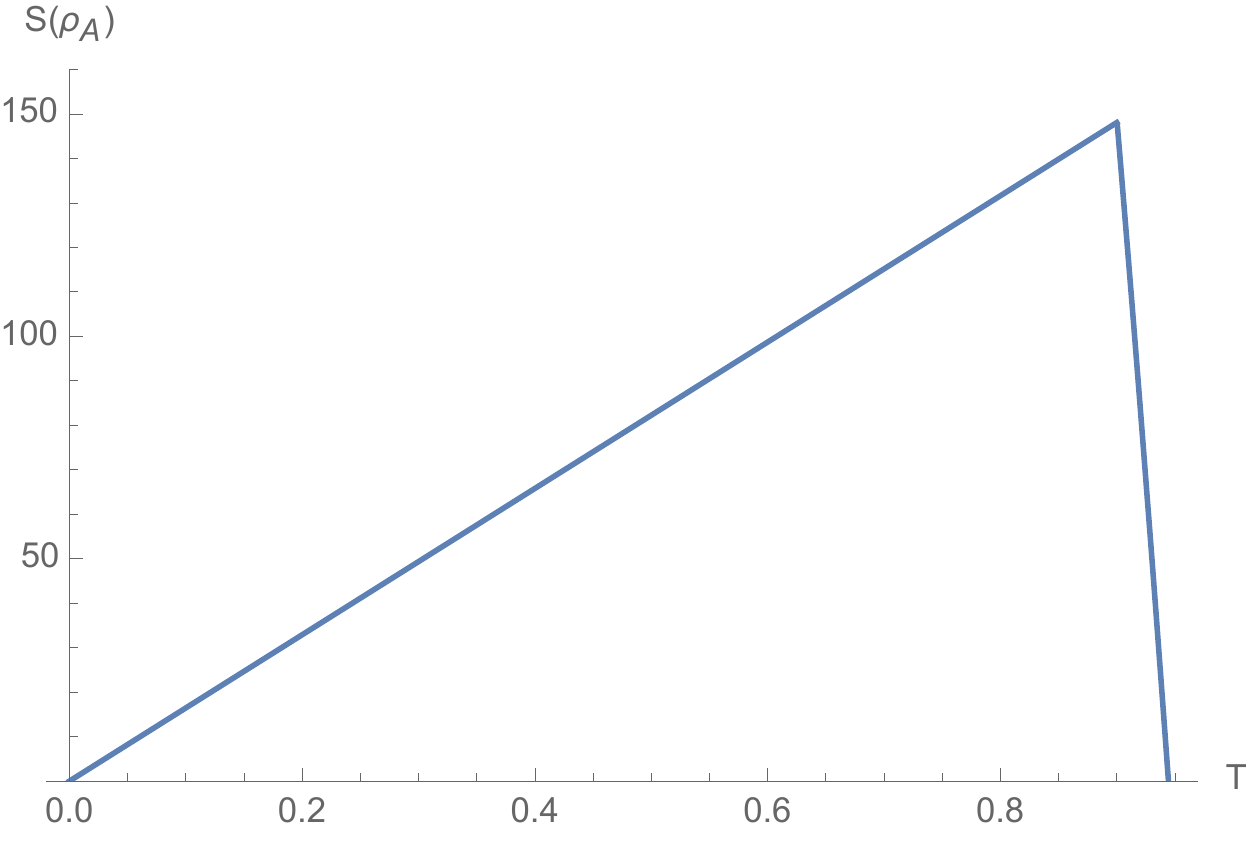}
  \hspace{0.5cm}
    \caption{\small{ {\bf Left} :Plot of the generalized entropy $S_{{\rm gen}} (x)$  as a function of the size of the island in the interior.  {\bf Right}: The resulting Page curve as a function of the entanglement temperature $T=1/\beta$. Here we set the parameters to be $\phi_0=1700,\; \Lambda=500,\; c=50$ in both figures and $\beta=1$ for left figure.}}
    \label{fig:SgenAndPage}
\end{figure}

Now, let us compute the entanglement entropy $S(\rho_{A})$ of the universe $A$ through the island formula \eqref{eq:defent} with \eqref{eq:generalizedentropy}. 
For this purpose we need to extremaize the generalized  entropy for all possible intervals $\bar{C}$ whose end points are identified with quantum extremal surfaces.    In the  calculation, it is reasonable to assume that $\bar{C}$ is  on the reflection symmetric slice  $x^{+} =x^{-}$ , and is given by the union of two intervals $\bar{C} =\bar{C}_{1} \cup \bar{C}_{2}$  , $\bar{C}_{1}: -\f{\pi}{2}  < x^{+} \leq - \f{\pi x}{2} $, $\bar{C}_{2} :  \f{\pi x}{2} \leq x^{+} <\f{\pi}{2}$ with $0<x<1$. The generalize entropy reduces to a function of single variable $x$,
\be
S_{{\rm gen}} (x) = 2\Phi_{\beta} (x) +\f{2c}{3} \log \left[\f{\beta}{\pi \ve_{\rm{UV}}} \sinh \f{\pi^{2}}{2\beta} (1-x)\right] -\f{2c}{3} \log \left[\f{1}{\ve_{\rm{UV}}}  \sin \frac{\pi}{2} (1-x )\right], 
\ee
where $\ve_{\rm{UV}}$ is the UV cutoff.
We give a plot of the above function in the left panel of figure \ref{fig:SgenAndPage}.

\begin{figure}[t]
    \centering
    \includegraphics[width=8cm]{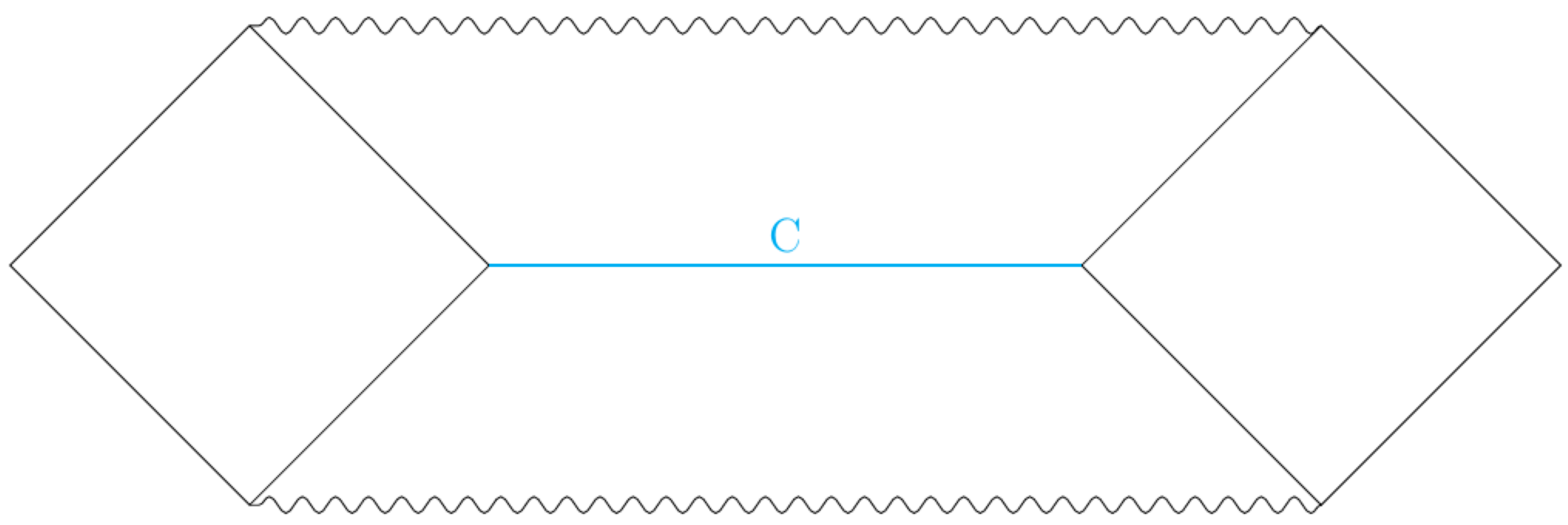}
    \hspace{0.5cm}

  \hspace{0.5cm}
    \caption{\small{ The location of the island $C$ in the black hole with the back-reaction, denoted by the blue line. }}
    \label{fig:island}
\end{figure}

In the $\beta \rightarrow 0$ limit, the quantum extremal surfaces almost coincide with the  classical bifurcation surfaces of the black hole.    This is because   the QESs approach  the spatial infinity and therefore the CFT entropy part  in $S_{{\rm gen}} (x) $  is vanishing  in this limit.   As a result, the island is identified with the causal shadow region in the black hole  interior (figure \ref{fig:island}).

By combining these result,  we get the following approximate expression for the entanglement entropy $S(\rho_{A})$,
\be
S(\rho_{A}) =
\begin{cases}
 S_{{\rm no-island}} = \f{\pi^{2}c}{3\beta}  & \beta> \beta_{c}\\[+10pt]
  S_{{\rm island}}= 2\left[\phi_{0}-\f{(\pi X_{\beta})^{2}}{\Lambda} \right]
   &\beta <\beta_{c},
\end{cases}
\label{eq:entwithoutshock}
\ee
where $\beta_{c}$ is the critical inverse temperature satisfying $ S_{{\rm no-island}}= S_{{\rm island}} $. We plot the Page curve by using the above expression in the right panel of figure \ref{fig:SgenAndPage}.

\section{Black hole interior in the presence of shock wave}
\label{section:BHshock}

We have seen,  by making the entanglement between two universes stronger, the size of the black hole interior in the gravitating universe $B$ gets larger. So in some sense this interior region is created by the entanglement between the degrees of freedom in the gravitating universe $B$ and the ones in the other (non-gravitating) universe. To sharpen the intuition, in this section,  we would like to ask how do the local operations in the gravitating universe B change the entanglement.

We imagine an experimental physicist in  a lab has this system $AB$  of two disjoint universes (or the larger Minkowski space containing AB  as in figure \ref{fig:SetupMin}), and can perform any local operation on the gravitating universe B even in the black hole interior. Such local operation can be modeled by a shock wave in the null directions  along which  the  CFT stress energy tensor  has a delta functional peak. Such a peak of the stress energy tensor can back-react to the dilaton profile through the equations of motion \eqref{eq: EOMs}.  

We start from the state  $|\Psi \ra$ on AB,  prepared by inserting a local operator $\mathcal{O}_{B}$ in the gravitating universe B to the thermofield double state
\be 
 |\Psi \ra =(1_{A} \otimes  \mathcal{O}_{B})   | TFD\ra= \f{1}{\s{Z(\beta)}}\sum_{i=0}^{\infty}  e^{-\f{\beta}{2} E_{i}}\;  | i \ra_{A} \; \otimes  \mathcal{O}_{B} \;  | \psi_{i} \ra_{B}\;. \label{eq:exstate}
\ee

We are  interested in its  entanglement entropy $S(\rho_{A})$ of the above state, which is  computed by the island formula \eqref{eq:defent}. Since $S_{{\rm no-island }}$ does not change by the insertion of $\mathcal{O}_{B}$,  we focus on $S_{{\rm island }}$ given by the  generalized entropy,
\be
S_{{\rm gen}}= {\rm Ext}_{\bar{C}} \;\left[  \Phi(\p \bar{C}) + S_{\beta, E}[\bar{C}] -S_{{\rm vac}} [\bar{C}] \right],
\label{eq;sgen}
\ee
where  $S_{\beta, E}[\bar{C}]$ denotes the   CFT entanglement entropy in the presence of the shock wave created by  $\mathcal{O}_{B}$.

Let $(x^{+}_{0},x^{-}_{0})$  be the location of the insertion of the operator $\mathcal{O}$. Then the reduced density matrix of the  universe B is 
\be
\rho={\rm tr}_{A} |\Psi \ra \la \Psi| =\f{1}{Z_{\mathcal{O}}}\; e^{-\ve H} \mathcal{O}(x^{+}_{0},x^{-}_{0})\rho_{\beta}\;\mathcal{O}^{\dagger}(x^{+}_{0},x^{-}_{0})\;e^{-\ve H}.
\label{eq:denmat}
\ee
Here we introduced a UV regulator $\ve$, to make the density matrix normalizable, and the normalization factor
\be
Z_{\mathcal{O}} = \la \mathcal{O}(2i \ve) \mathcal{O}(0) \ra_{\beta}, 
\ee
which ensures ${\rm tr} \; \rho=1$. We also denote,  $\la \cdots \ra_{\beta} \equiv {\rm tr} [\rho_{\beta} \cdots]$.

This local operator $\mathcal{O}$ affects the stress energy tensor expectation value, and therefore the dilaton profile.   The stress energy tensor expectation value can be computed by the three point functions 
${\rm tr} [\rho_{\beta} T_{\pm \pm} \mathcal{O}\mathcal{O}]$, and it reads
\be 
\la \Psi | T_{++} (x^{+}) | \Psi \ra = \f{c}{24\pi} \left(\f{2\pi}{\beta} \right)^{2} +E_{{\rm Shock}}\; \delta (x^{+}- x_{0}^{+}), \quad  \la \Psi | T_{--} (x^{-}) | \Psi \ra = \f{c}{24\pi} \left(\f{2\pi}{\beta} \right)^{2} +E_{{\rm Shock}}\; \delta (x^{-}- x_{0}^{-}),
\label{eq;shockstress}
\ee
in the $\ve \rightarrow 0$ limit.  The coefficient of the delta functions is related to the conformal dimension $\Delta$ of this local operator
\be
E_{{\rm Shock}}= \f{\Delta}{\ve}.
\ee
Therefore, the  insertion of a local operator creates a pair of shock waves in the black hole geometry, one is left moving and the other is right moving. The existence of these shocks is manifested by the delta functional peaks of the CFT stress energy tensor expectation value. For simplicity, we  write $E \equiv E_{{\rm Shock}}$ below.

\subsection{Dilaton part}

Let us discuss in detail how the shock wave changes the dilaton $\Phi$. 
It satisfies the following equations of motion, 
\be 
\begin{split}
	-e^{2 \omega} \p_{\pm}\left(e^{-2 \omega} \p_{\pm} \Phi\right) &= 2 X_{\beta}+E\, \delta\left(x^{\pm}-x_{0}^{\pm}\right), \\[+10 pt]
\p_{+} \partial_{-} \Phi &= \f{\Lambda}{4}  e^{2 \omega}.
\end{split}
\ee

 These equations are obtained by substituting the stress tensor expectation value  \eqref{eq;shockstress} to \eqref{eq: EOMs} for arbitrary $\la T_{\pm \pm} \ra$.  By solving these equations, we obtain, the dilaton profile in the presence of shock wave,
\be 
\begin{split}
\Phi &= \phi_{0} + \f{\Lambda}{4} \tan x^{+}\tan x^{-} -X_{\beta} (x^{+} \tan x^{+} + x^{-} \tan x^{-} )\\
&-E \cos^{2} x^{+}_{0} (\tan x^{+}-\tan x^{+}_{0})\theta(x^{+} -x^{+}_{0})-E \cos^{2} x^{-}_{0} (\tan x^{-}-\tan x^{-}_{0})\theta(x^{-} -x^{-}_{0}),
\end{split}
\label{eq:completedl}
\ee
where $\theta(x)$ is the step function, 
\be
\theta (x) =
\begin{cases}
    1 &x>0 \\
    0&x<0.
\end{cases}
\ee

\subsection{Classical extremal surfaces}\label{subsec:Classical extremal surfaces}

\begin{figure}[t]
    \centering
    \includegraphics[width=5cm]{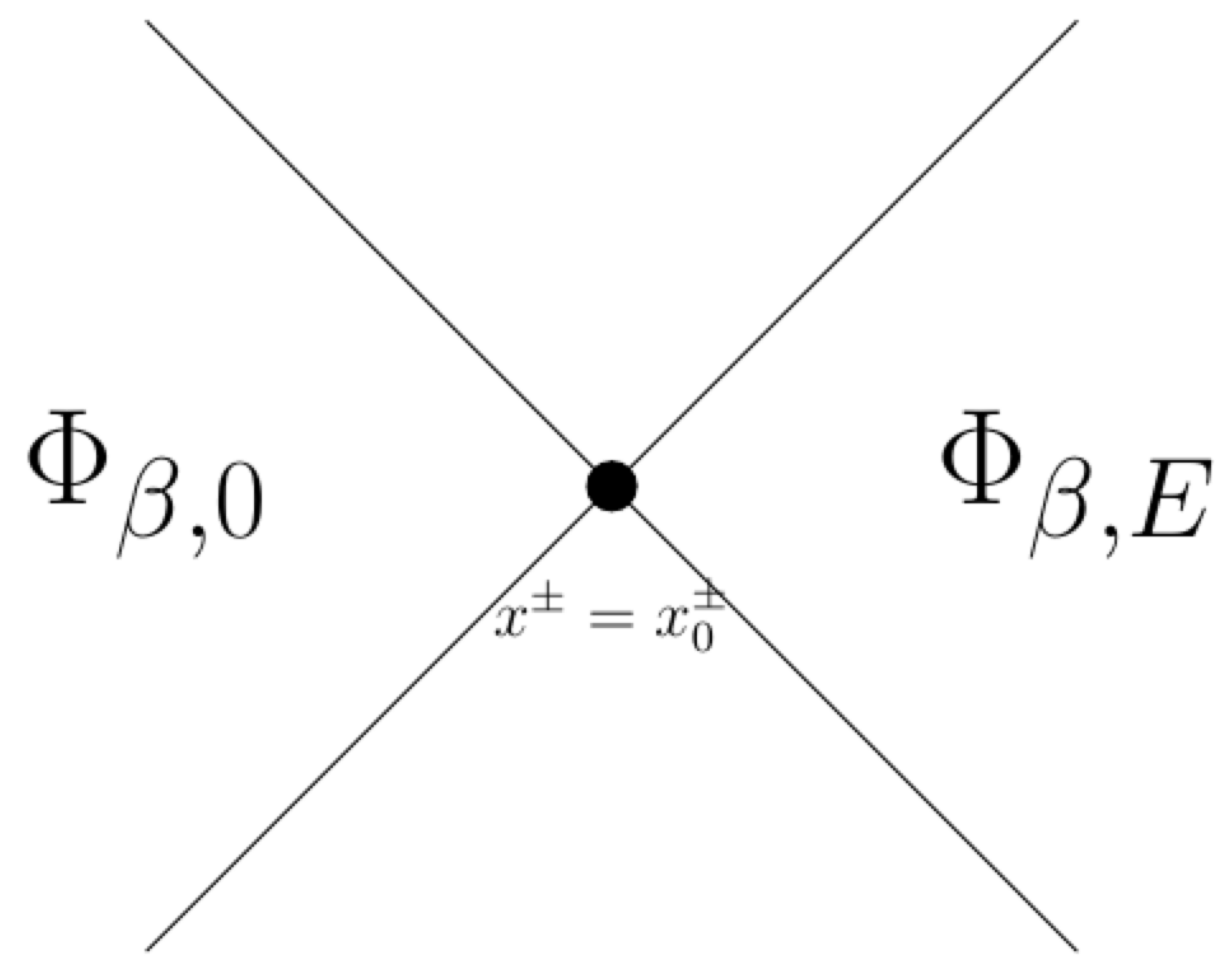}
    \hspace{0.5cm}

  \hspace{0.5cm}
    \caption{\small{ The dilaton profile \eqref{eq:completedl} in the presence of the shock wave. In the right wedge of the local operator,  $x^{\pm} >x^{\pm}_{0}$, we have $\Phi= \Phi_{\beta, E}$ with \eqref{eq:phibetaE}. On the left wedge, $x^{\pm} <x^{\pm}_{0}$, the dilaton profile coincides with $\Phi_{\beta,0}$, which is identical to \eqref{eq:dilbeta} }.}
    \label{fig:classification}
\end{figure}

Now we would like to specify the classical  extremal surfaces in the spacetime with the dilaton profile  \eqref{eq:completedl}. We will see that the actual locations of these surfaces highly depend on where we insert the local operator. In the right wedge of the local operator $x^{\pm} >x^{\pm}_{0}$, the dilaton  coincides with $ \Phi_{\beta,E}$ defined by 
\be
 \begin{split}
\Phi_{\beta,E} &= \phi_{0} + \f{\Lambda}{4} \tan x^{+}\tan x^{-} -X_{\beta} (x^{+} \tan x^{+} + x^{-} \tan x^{-} )\\
&-E \cos^{2} x^{+}_{0} (\tan x^{+}-\tan x^{+}_{0})-E \cos^{2} x^{-}_{0} (\tan x^{-}-\tan x^{-}_{0}).
\end{split}
\label{eq:phibetaE}
 \ee
 In  the left wedge  $  x^{\pm}<x^{\pm}_{0}$, it agrees with with the original profile $\Phi  =  \Phi_{\beta,E=0} \equiv \Phi_{\beta}$ \eqref{eq:dilbeta}, see figure \ref{fig:classification}.  We also argued that in the absence of the shock wave, ie, $E=0$,  the black hole has a causal shadow region in its interior, so it has two bifurcation surfaces. In the presence of the shock,
  the dilaton profile \eqref{eq:completedl} has also two  critical points, one is near the left spatial infinity $ (x^{+}, x^{-}) = (-\f{\pi}{2},-\f{\pi}{2})$ and the other is near the right spatial infinity $ (x^{+}, x^{-}) = (\f{\pi}{2},\f{\pi}{2})$. In this section, we only consider the operator insertions, which do not change the location of the left horizon of the undeformed dilaton $\Phi_{\beta}$. This is equivalent to restrict the range of the insertion to $0<x^{+}_{0} +x^{-}_{0}$.  Under this restriction, we can  focus on the change of the location of the right critical point below.  The discussion  for operator insertions in the $ x^{+}_{0} +x^{-}_{0} <0$  region can be made similarly.
 
 To identify the right extremal surface, it is convenient to introduce two characteristic points of the dilaton profile $\Phi$.  First, let $x^{\pm}=x^{\pm}_{H} (0) $ be the critical point of  the original dilaton profile $\Phi_{\beta} (x^{\pm})$ ,ie
 \be
 \p_{\pm}\Phi_{\beta}|_{x^{\pm}=x^{\pm}_{H} (0)} =0 \rightarrow \tan \; x^{\pm}_{H}(0)  =\f{2\pi}{\Lambda} X_{\beta}.\label{eq:originalhorizon}
 \ee

The second characteristic point is  $x^{\pm}=x^{\pm}_{H} (E) $, which is the critical point of the deformed dilaton profile  $\Phi_{\beta, E}$ \eqref{eq:phibetaE}.  It satisfies
 \be
 \tan \; x^{\pm}_{H}(E) =\f{4}{\Lambda} \left( \f{\pi}{2}X_{\beta} + E\cos^{2} x_{0}\right). \label{eq:deformedhorizon}
 \ee

 These are candidates of the extremal surfaces of the total dilaton \eqref{eq:completedl}, but  whether  these are  the critical points of the actual dilaton \eqref{eq:completedl}  depends on the location of the local operator  $x^{\pm}= x^{\pm}_{0}$. 

In order to simplify the discussion below, instead of exhausting all possible cases,  let us first consider the  symmetric  insertions $x^{+}_{0} =x^{-}_{0} \equiv x_{0}$.  Now these two candidate extremal  surfaces are also symmetric, ie they are on the reflection symmetric slice : $x^{+}_{H} (0) =x^{-}_{H} (0)\equiv   x_{H} (0)$, and $x^{+}_{H} (E) =x^{-}_{H} (E)  \equiv   x_{H} (E)$. In general, the relation  $x_{H}(0)< x_{H}(E)$ holds. 
In this setup, there are three distinct cases for the operator insertions (see figure \ref{fig:classification}). Namely, the location of the operator is (1) behind the original horizon $x_{0} < x_{H}(0)\;$,  (2) in  the middle of two horizons,  $ x_{H}(0)<x_{0} <x_{H}(E)\;$ and (3) in the  exterior of the deformed horizon  $x_{H}(E)<x_{0}$.

\subsubsection*{Case 1}
In the first case, the local operator $\mathcal{O}$ is inserted to the left of the original horizon $:x_{0} < x_{H}(0)$ (left panel of figure 
\ref{fig:thereecase}). In this case,  only the extremal surface is the deformed horizon $x^{\pm}= x_{H} (E)$. This is because  $x^{\pm} = x_{H}(0)$  is not a critical point of the dilaton profile \eqref{eq:completedl}, since around this point, this profile  conicides with  the deformed one $\Phi_{\beta, E}$ \eqref{eq:phibetaE}, due to the condition  $x_{0} < x_{H}(0)$. 
The dilaton value at the extremal surface is given by $\Phi(x_{H} (E)) =\Phi_{\beta,E}(x_{H} (E))$.

\subsection*{Case 2}

In the second case, the local operator is inserted in between two would be  extremal surfaces $ x_{H}(0)<x_{0} <x_{H}(E)\;$ (middle panel of  figure 
\ref{fig:thereecase}).  In this case, both horizons $x^{\pm}=x^{\pm}_{H}(0)$ and  $x^{\pm}=x^{\pm}_{H}(E)$ are actually extremal surfaces of the dilaton profile \eqref{eq:completedl}.

\subsection*{Case 3} 

\begin{figure}[t]
    \centering
    \includegraphics[width=4.5cm]{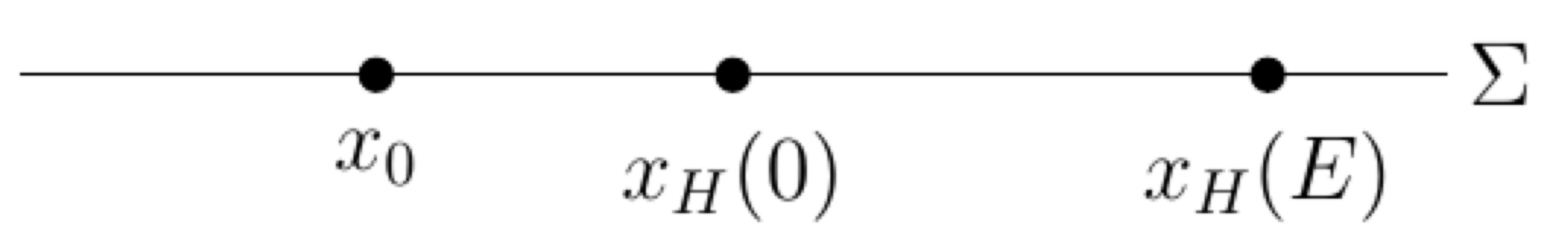}
    \hspace{0.5cm}
\includegraphics[width=4.5cm]{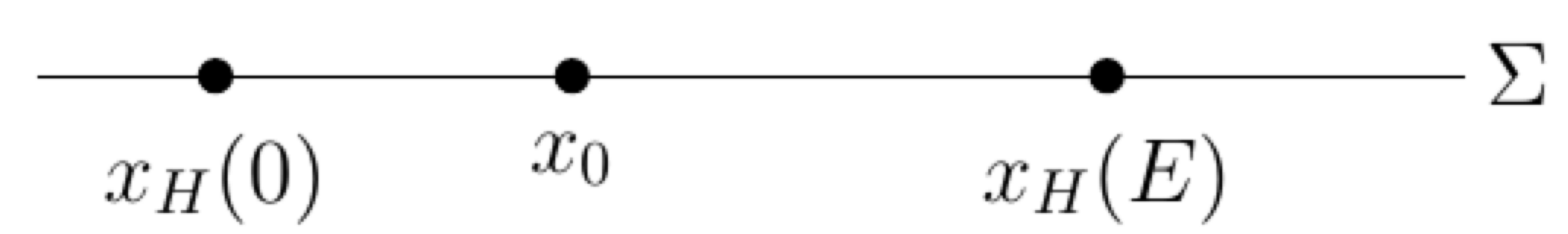}
  \hspace{0.5cm}
\includegraphics[width=4.5cm]{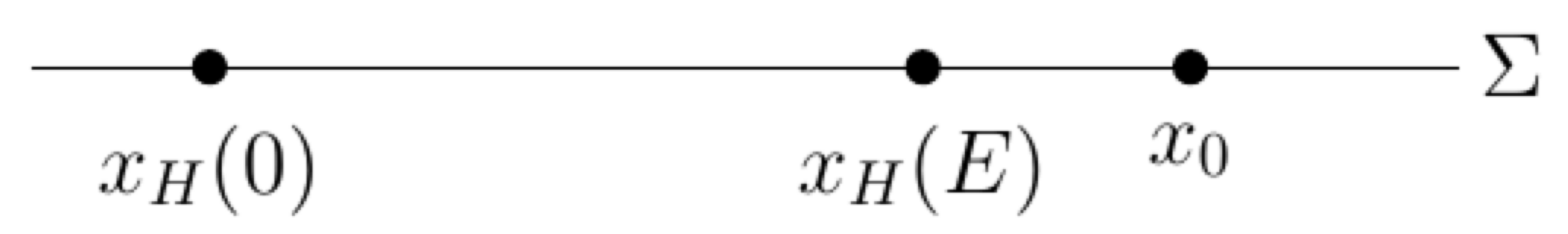}
    \caption{\small{Three possible locations of the local operator $\mathcal{O}$ on the reflection symmetric slice $\Sigma: x^{+}=x^{-}$. 
    {\bf Left }: When $x_{0} <x_{H}(0)< x_{H}(E)$, the extremal surface is at $x^{\pm} =x_{H}(E)$.  {\bf Middle} : When $ x_{H}(0)<x_{0}< x_{H}(E)$, both $x^{\pm} =x_{H}(0)$ and $x^{\pm} =x_{H}(E)$ are extremal. {\bf Right}: When $x_{H}(0)< x_{H}(E)< x_{0}$, the extremal surface is at $x^{\pm} =x_{H}(0)$. }}
    \label{fig:thereecase}
\end{figure}

In the third case,  the local operator is inserted to the right of the deformed horizon $x_{H}(E)<x_{0}$ (right panel of  figure 
\ref{fig:thereecase}) .
In this case, only the extremal surface is the original horizon $x^{\pm}=x_{H}^{\pm}(0)$. This is again because the deformed  horizon 
$x^{\pm}=x_{H}^{\pm}(E)$ is not the critical point of the dilaton. 
The dilaton value at the  extremal surface  is  given by $\Phi_{\beta}(x_{H}^{\pm}(0))$.

\subsection{CFT entropy part}

The second ingredient of the generalized entropy \eqref{eq;sgen} is the CFT entropy  $S_{\beta, E}[\bar{C}]$  of the density matrix \eqref{eq:denmat}  on $ \bar{C} $ in the gravitating universe $B$.  
Because we focus on the high temperature limit $\beta \rightarrow 0$, this region is the  disjoint union of  two pieces  $\bar{C}=\bar{C}_1\cup\bar{C}_2 $, as  in the previous section. 
We put the coordinates of $ \bar{C} $ as follows;
\begin{equation*}
	\left\{\begin{array}{l}
		x_{2}^{\pm}=-\frac{\pi}{2},\\
		x_{3}^{\pm}=x_{3} \pm t_{3},
	\end{array}\right. \text { for } \bar{C}_{1}
\end{equation*}
\begin{equation*}
	\left\{\begin{array}{l}
		x_{5}^{\pm}=x_{5} \pm t_{5}, \\
		x_{6}^{\pm}=\frac{\pi}{2}.
	\end{array}\right. \text { for } \bar{C}_{2}
	\label{eq:C2}
\end{equation*}
In the $\beta \rightarrow 0$ limit, $x^{\pm}_{3} \rightarrow x^{\pm}_{2}$  and $x^{\pm}_{5} \rightarrow x^{\pm}_{6} $ holds.
Therefore, in the absence of the shock wave $E=0$,  the CFT entanglement  entropy for $\bar{C}=\bar{C}_1\cup\bar{C}_2 $ at finite temperature $ \beta$ is given by 
\begin{equation}
	\begin{split}
		S_{\beta}[\bar{C}]=&
		\frac{c}{6} \log \left[\frac{\beta}{\pi \varepsilon_{UV}} \sinh \left(\frac{\pi}{\beta}\left(x_{3}^{+}-x_{2}^{+}\right)\right)\right] +\frac{c}{6} \log \left[\frac{\beta}{\pi \varepsilon_{UV}} \sinh \left(\frac{\pi}{\beta}\left(x_{3}^{-}-x_{2}^{-}\right)\right)\right]\\
		&+\frac{c}{6} \log \left[\frac{\beta}{\pi \varepsilon_{UV}} \sinh \left(\frac{\pi}{\beta}\left(x_{6}^{+}-x_{5}^+\right)\right)\right] +\frac{c}{6} \log \left[\frac{\beta}{\pi \varepsilon_{UV}} \sinh \left(\frac{\pi}{\beta}\left(x_{6}^{-}-x_{5}^{-}\right)\right)\right].
	\end{split}
\end{equation} 
We also need the CFT  entropy for $\bar{C}=\bar{C}_1\cup\bar{C}_2 $ at zero temperature, and it is given by
\begin{equation}
	\begin{split}
		S_{{\rm vac}}[\bar{C}]=\frac{c}{6} & \log \left[\frac{1}{\varepsilon_{U V}} \sin \left(x_{3}^{+}-x_{2}^{+}\right)\right]+\frac{c}{6} \log \left[\frac{1}{\varepsilon_{U V}} \sin \left(x_{3}^{-}-x_{2}^{-}\right)\right] \\
		+& \frac{c}{6} \log \left[\frac{1}{\varepsilon_{U V}} \sin \left(x_{6}^{+}-x_{5}^{+}\right)\right] +\frac{c}{6} \log \left[\frac{1}{\varepsilon_{U V}} \sin \left(x_{6}^{-}-x_{5}^{-}\right)\right].
	\end{split}
\end{equation}

\subsubsection{CFT entropy for a single interval} \label{subsubsection:CFTentropy}

Next, we discuss  the entanglement  entropy $S_{\beta,E}[\bar{C}]$ in the presence of a shock wave. This kind of  entanglement  entropy was studied in \cite{Caputa:2015waa}, which we review in Appendix  \ref{section:EEderivation}.
As a warm-up, let us compute the  entanglement entropy  $S_{\beta,E}[\bar{C}] $ of the single interval,
\be
\bar{C}: x_{5}^{\pm} <x^{\pm} < x^{\pm}_{6}= \f{\pi}{2}.
\ee
which ends at the asymptotic infinity $x^{\pm}_{6} =\f{\pi}{2}$.
  In presenting the expression of the CFT entropy, it is convenient to first fix the subsystem $\bar{C}$ ie, fixing $x_{5}^{\pm}$ .

This entanglement entropy  can be computed, first writing the R\`enyi entropy ${\rm tr} \rho_{\bar{C}}^{n}$ by a four point function involving twist operators, 
\be
{\rm tr} \rho_{\bar{C}}^{n} = {\rm tr}  \left[ \rho_{\beta} \; \mathcal{O}^{\otimes n} (x_{1})\sigma_{n} (x_{5})\sigma_{-n} (x_{6})   \mathcal{O}^{\otimes n}  (x_{4}) \right],
\ee
and taking $n \rightarrow 1$ limit of the correlator. Here $x_1$ and $x_4$ are related to $x_0$ through \eqref{eq:x1x4andx0}.
When the central charge of the theory is large $c \gg 1$, and its spectrum is sparse, the right hand side can be  approximated by the vacuum conformal block with a choice of branch \cite{Caputa:2015waa}. Again details can be found  in the Appendix.

\begin{figure}[t]
    \centering
    \includegraphics[width=5cm]{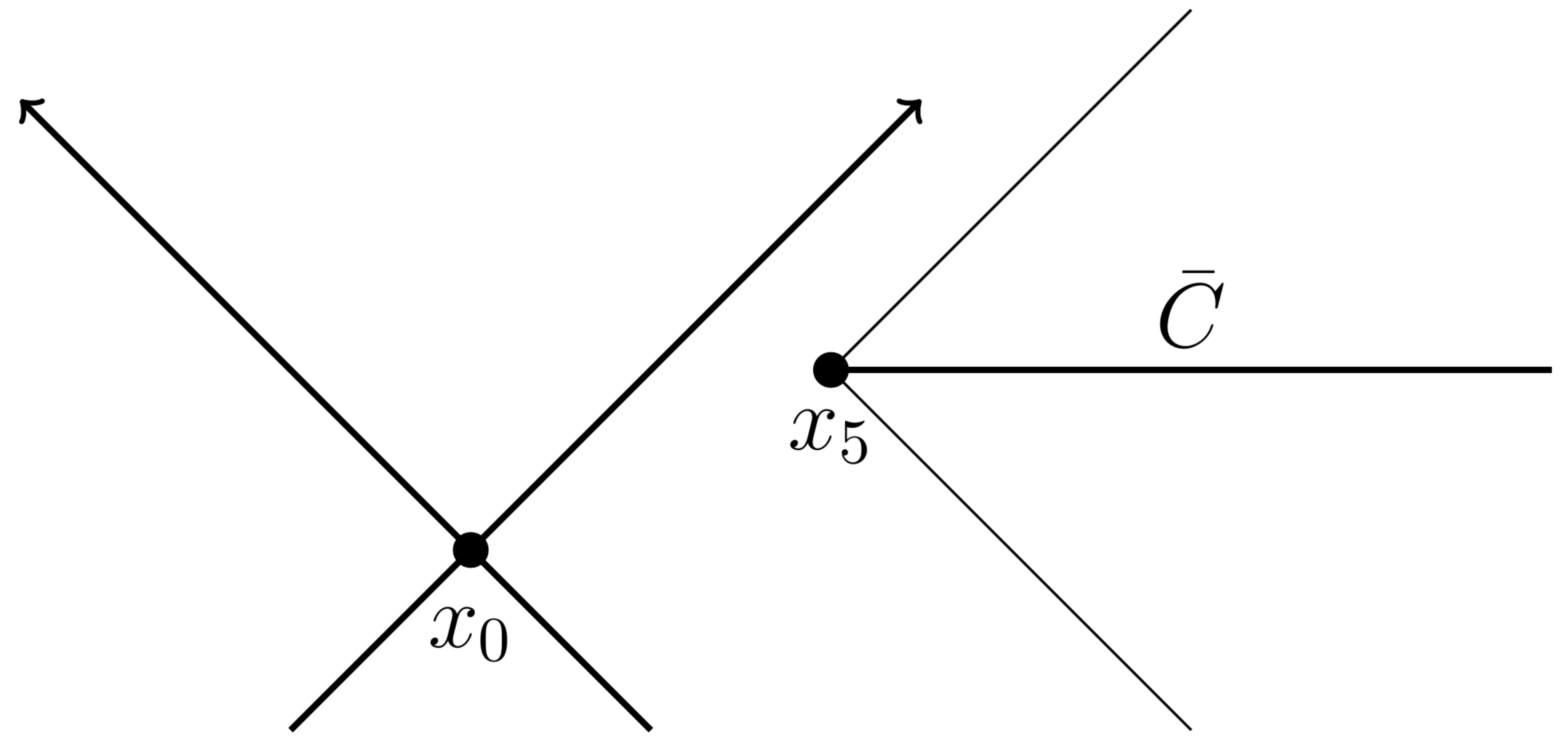}
    \hspace{0.5cm}
\includegraphics[width=3.5cm]{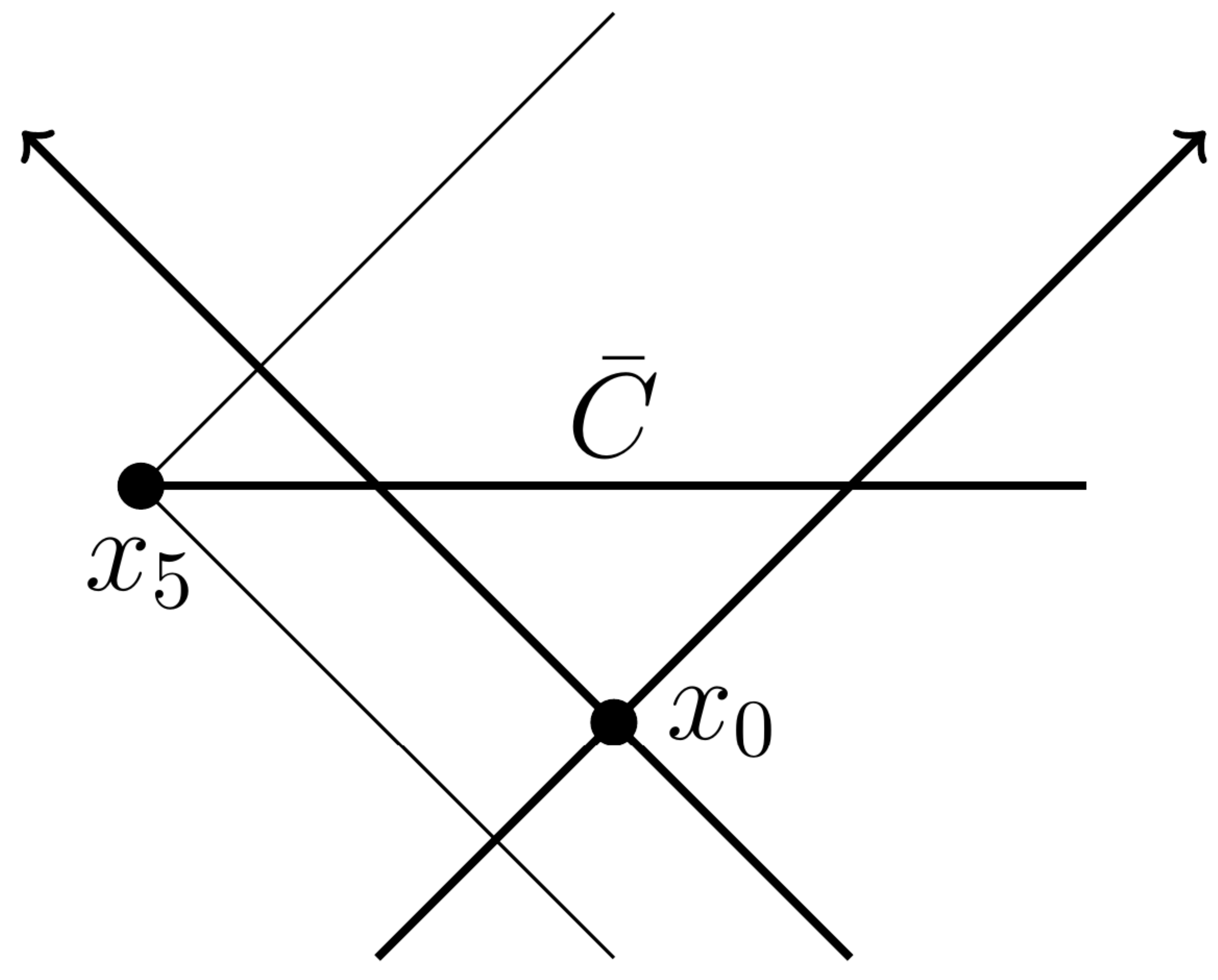}
  \hspace{0.5cm}
\includegraphics[width=4.5cm]{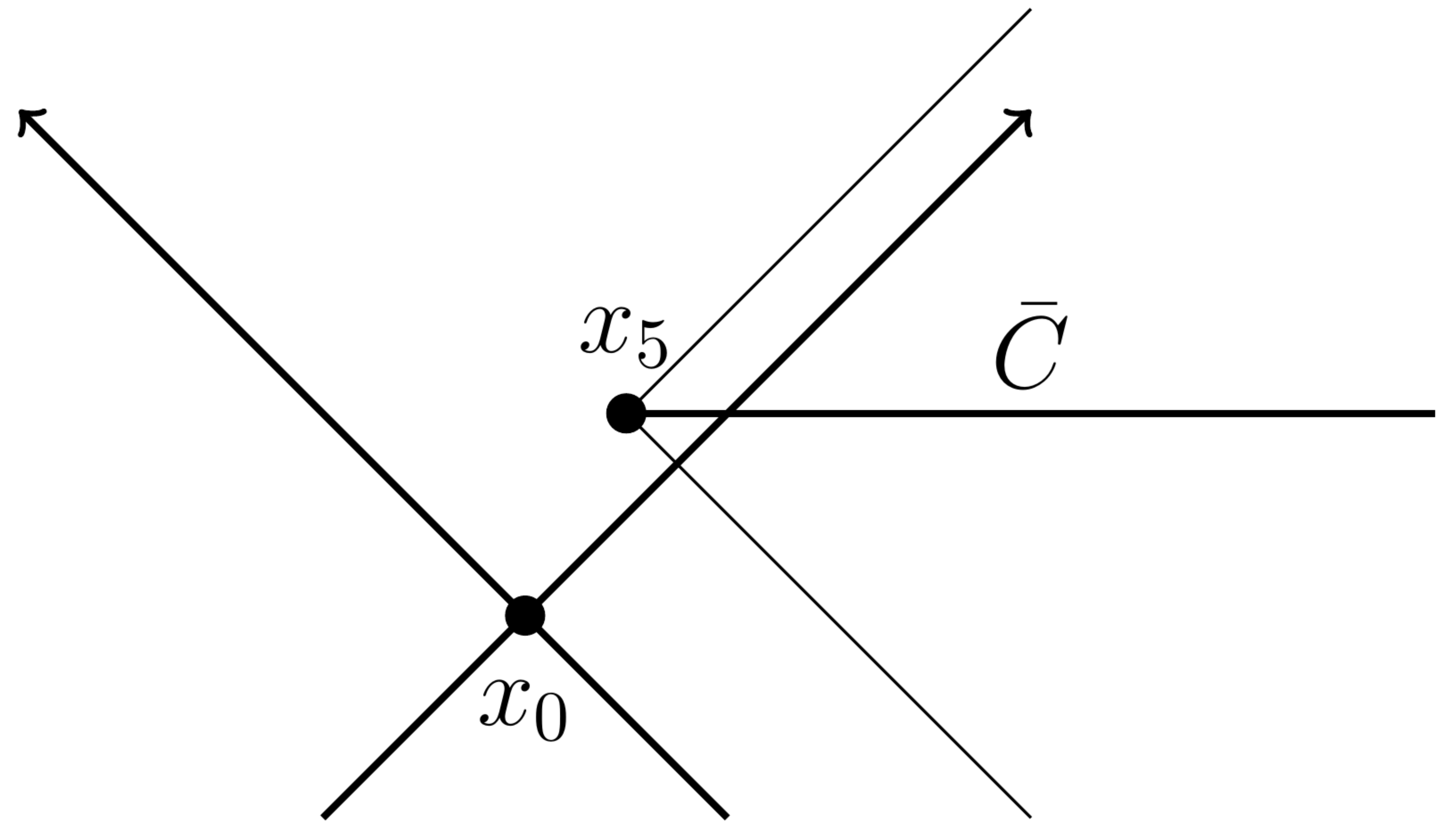}
    \caption{\small{ 
    {Three possible causal relations between the operator  at $x^{\pm} =x^{\pm}_{0}$ and the interval $\bar{C}$. \bf Left } and {\bf Middle} : The local operator is spatially separated from $\bar{C}$. In this case, the non-trivial part of the CFT entropy $\Delta S$ in the generalized entropy  is vanishing due to causality. {\bf Right}: When the local operator and the $\bar{C}$ are causally connected, the right mover emitted by the quench can enter the causal diamond of $D[\bar{C}]$. Only in this case, $\Delta S$ is non vanishing. }}
    \label{fig:causalrelations}
\end{figure}

The possible form of the CFT entropy is constrained by the  causal relation  between the location of the  operator  $x^{\pm} =x^{\pm}_{0}$ and the region $\bar{C}$ \cite{Calabrese:2007mtj,Caputa:2014vaa,Nozaki:2013wia,Asplund:2014coa,Caputa:2015waa, Ugajin:2013xxa,Asplund:2013zba,Nozaki:2014hna,Nozaki:2014uaa,Caputa:2014eta,David:2016pzn,Hartman:2015lfa}. Indeed, the  insertion creates shock waves, both left moving and right moving,   which is roughly speaking interpreted  as an entangled pair of particles.  Then  the CFT entropy  can be non-trivial only when one of these shock waves  enters the causal diamond of the region $D[\bar{C}]$,   whereas its partner does not.  Therefore, for the fixed end point of the  subsystem, $x^{\pm} =x^{\pm}_{5}$,  we have four possible behaviors of the entropy, according to the location of the local operator  relative to the end point as in figure \ref{fig:causalrelations}.

(1) When the end point $x_{5}$ is  in the right wedge of  the  location  of the operator  $x_{0}$,  ie, $x^{\pm}_{5}  >x^{\pm}_{0}$, both left mover and  right mover do not enter the causal diamond $D[\bar{C}]$ of the subsystem $\bar{C}$. Therefore, the shock wave can not affect  the state on the subsystem  $\bar{C}$,   the entanglement entropy  remains unchanged  due to causality. 

(2) Similarly when   the end point  is in the  left wedge of  the insertion  $x^{\pm}_{0}  >x^{\pm}_{5}$,  both shock waves enter the causal  diamond,  and the entanglement entropy again remains unchanged.

(3) When the local operator  is in the future or past  of the end point,  the entanglement entropy can be non-trivial.  When it is in the future light cone of the end point $x_{5}$,  ie $x_{5}^{-}>x_{0}^{-}$ and $x_{0}^{+} > x_{5}^{+}$, then only the left moving shock contributes to the entropy.  The difference between this entanglement entropy and  the thermal one  $\Delta S \equiv S_{\beta, E}[\bar{C}] -S_{\beta}[\bar{C}]$ is  given by \cite{Caputa:2015waa}
\be
\Delta S_{F}=\f{c}{6} \log \left[\f{\beta}{\pi \ve} \f{\sin \pi \alpha}{\alpha} \f{\sinh \f{\pi}{\beta} (x^{+}_{0}-x^{+}_{5})\sinh \f{\pi}{\beta} (x^{+}_{6}-x^{+}_{0})}{\sinh \f{\pi}{\beta} (x^{+}_{6}-x^{+}_{5})} \right].
\label{eq:futureentropy}
\ee

When the operator is in the past light cone of the end point, $x_{0}^{-}>x_{5}^{-}$ and $x_{5}^{+} > x_{0}^{+}$, then
only the right mover contributes, and the entropy difference is
\be
\Delta S_{P}=\f{c}{6} \log \left[\f{\beta}{\pi \ve} \f{\sin \pi \alpha}{\alpha} \f{\sinh \f{\pi}{\beta} (x^{-}_{0}-x^{-}_{5})\sinh \f{\pi}{\beta} (x^{-}_{6}-x^{-}_{0})}{\sinh \f{\pi}{\beta} (x^{-}_{6}-x^{-}_{5})} \right].
\label{eq:shockentropy}
\ee

\subsubsection{CFT entropy for two disjoint intervals}
\label{subsubsection:CFTdisjoint}

In the  actual calculations of the generalized entropy, we need an expression of the CFT entanglement entropy for two disjoint intervals $\bar{C} =\bar{C}_{1} \cup \bar{C}_{2}$.  Again,  the behavior of the entanglement entropy is strongly constrained by causality. In the previous section,  we saw that  these two intervals become smaller and smaller $ \bar{C}_{1}, \; \bar{C}_{2}\rightarrow 0$ in the high temperature limit $\beta \rightarrow 0$. This means that the entropy of this two disjoint intervals gets factorized,
\be
S[\bar{C}] = S[\bar{C_{1}}] +S[\bar{C_{2}}].
\ee
so the result for  the single interval is enough to fix the generalized entropy in this limit.
For simplicity,  we assume the shock wave does not intersect the left interval $\bar{C}_{1}$, ie $x^{\pm}_{0}$,  so only  the entropy of the right interval $S[\bar{C_{2}}]$ can change non-trivially.

\subsection{Quantum extremal surfaces}

We are interested in, how the dominant quantum extremal surface changes as we tune the location of the operator $x_{0}$. We are especially interested in the high temperature limit $\beta \rightarrow 0$, where the classical horizon is  getting close to the infinity $x^{\pm} =\f{\pi}{2}$. It is convenient to decompose the CFT entropy  $S_{\beta, E}[\bar{C}]- S_{{\rm vac}}[\bar{C}]$  in \eqref{eq;sgen} , into the trivial part  $S_{\beta}[\bar{C}] -S_{{\rm vac}}[\bar{C}] $ which does not involve the shock wave, and the non-trivial part $\Delta S =S_{\beta, E}[\bar{C}] -S_{\beta}[\bar{C}]$. Then the trivial part   does not play any role in the generalized entropy in the high temperature limit. As we saw  in the previous  section, the classical extremal surfaces in this   limit are given by the bifurcation surfaces  of the black hole.  Therefore, we can focus on the non-trivial part of the CFT entropy to find the quantum extremal surfaces. Without going into detail, let us  describe  two limiting cases. 

First,
when  the insertion is made in  the deep interior of the black hole $x_{0}  \sim 0$, the true quantum extremal surface almost coincides with the classical horizon of the deformed  black hole (The one with back-reaction of the shock wave) at $x^{\pm} =x^{\pm}_{H} (E)$, defined in \eqref{eq:deformedhorizon}.  This is because  the local  operator is spatially separated from the horizon, so  the non-trivial part of the CFT entropy $\Delta S$ is vanishing  due to causality. 

On the other hand, if the operator is inserted  at the exterior of  the horizon, ie $x_{0} > x_{H} (E)$, then, since again the non-trivial part of the CFT entropy is vanishing,   the QES coincides with the classical  extremal surface, which is identified with the horizon of the  original black hole (ie, black hole without the shock wave) at $x^{\pm} =x^{\pm}_{H} (0)$ \eqref{eq:originalhorizon}. Below, we  discuss  details of the  dynamics of the QESs.

We remark  that the generalized entropies also have the contribution from the the left classical extremal surface, which is independent of the location of the operator as long as the operator is inserted at the region $x_{0}^{+}+x_{0}^{-}>0$.
The contribution from the left extremal surface is given by \eqref{eq:dilatonvCES} in the high temperature limit and let us denote it by $S_{L}$.
Then, the generalized entropies for each case are given as follows.

\subsection*{Case 1}

When the local operator is inserted  inside of the original horizon (the left panel of figure \ref{fig:causalrelations}), $x_{0}<x_{H}(0)$, the quantum extremal surface coincides with the bifurcation surface of the deformed black hole at $x^{\pm} =x_{H}(E)$. Also,  since $\Delta S=0$ in this case, we get the following expression for the generalized entropy,
\begin{equation}
	\begin{aligned}
		S_{\mathrm{gen}}&=\Phi_{\beta,E}(x_{H}(E))+S_{L}\\
		&=\phi_{0}-\frac{4}{\Lambda}\left[\left(\frac{\pi X_\beta}{2}\right)^{2}+\pi X_\beta E \cos ^{2} x_{0}\right]
-\frac{4}{\Lambda} E^{2} \cos ^{4} x_{0}+2E\cos ^{2} x_{0} \tan x_{0}  +S_{L}.
	\end{aligned}
\end{equation}

\subsection*{Case 2}
When we insert the operator in between two bifurcation surfaces (the middle panel of figure \ref{fig:causalrelations})  , $x_{H}(0)<x_{0}<x_{H}(E)$, the generalized entropy is given by 
\be 
\begin{aligned}
S_{{\rm gen}} &={\rm min}\left\{\Phi_{\beta} (x_{H}(0)), \; \Phi_{\beta,E} (x_{H}(E))  \right\}+S_{L}\\[+10 pt]
		&=\phi_{0}-\frac{(\pi X_\beta)^2}{\Lambda} +\min \left\{0, -2E\cos^2 x_0 \left( \frac{2 \pi X_\beta}{\Lambda}  +\frac{2}{\Lambda} E \cos ^{2} x_{0} -\tan x_{0}\right) \right\}+S_{L}.
\end{aligned}\label{eq:originalGeneralizedE}
\ee
Again in this case the non-trivial part of the CFT entropy  $\Delta S$ vanishes.

The transition point $x^{\pm}= x_T $, at which the dominance in the minimization of the above expression changes,  satisfies the  equation
\begin{equation}
	\tan x_T=\frac{2 \pi X_\beta}{\Lambda}  +\frac{2}{\Lambda} E \cos ^{2} x_{T} .
\end{equation}

\subsection*{Case 3}
When the operator is inserted at the outside of the deformed horizon (the right panel of figure \ref{fig:causalrelations}), $x_{H}(E) <x_{0}$, we have
\begin{equation}
	\begin{aligned}
	S_{\text {gen }}&=\Phi_\beta\left(x_{H}(0)\right)+S_{L}\\
	&=\phi_{0}-\frac{(\pi X_\beta)^2}{\Lambda} +S_{L}.
\end{aligned}
\end{equation}

\subsection*{Net result}

\begin{figure}[t]
    \centering
    \includegraphics[width=9cm]{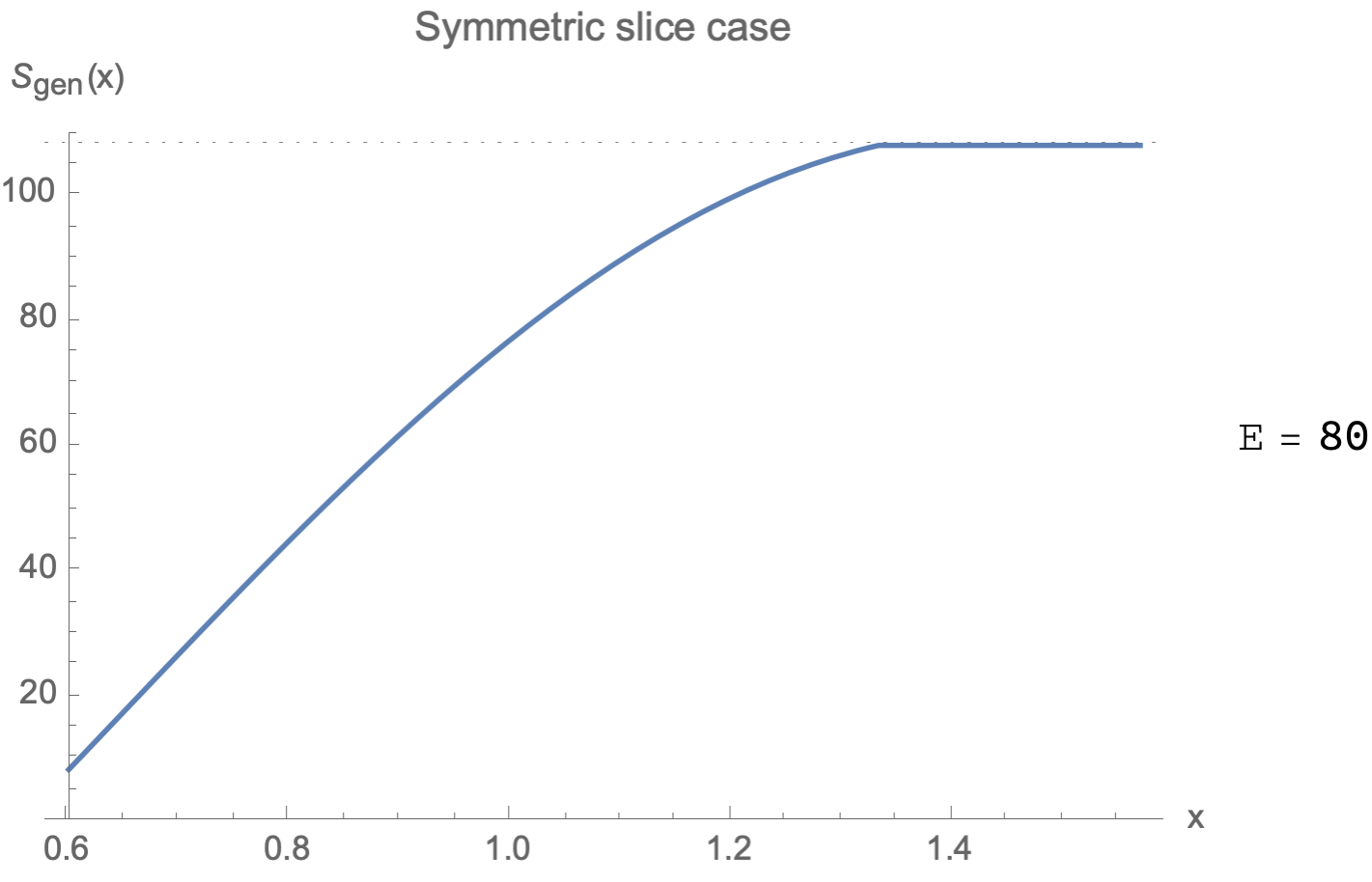}
    \hspace{0.5cm}
\includegraphics[width=9cm]{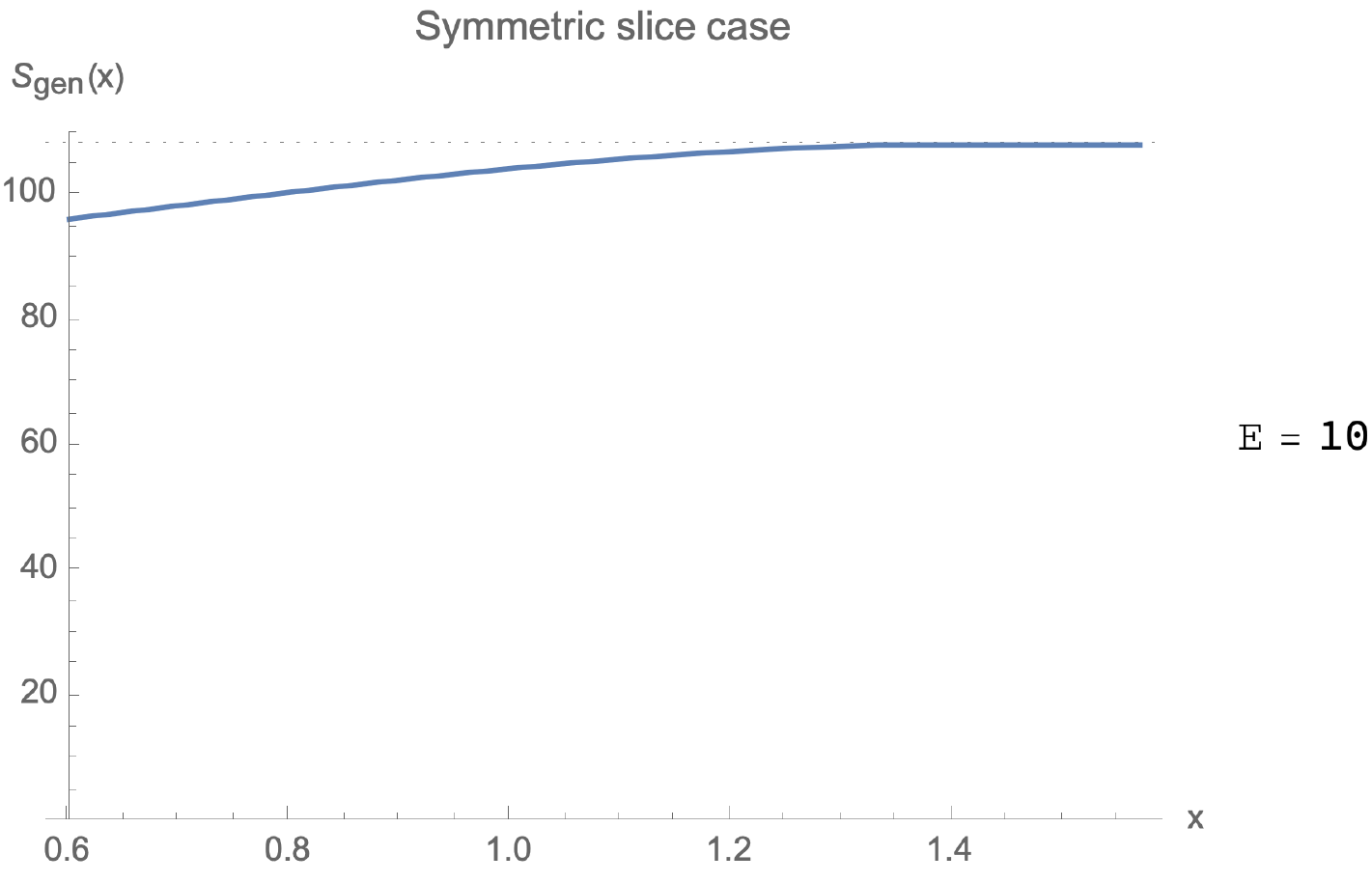}
  \hspace{0.5cm}
    \caption{\small{Plots of $ S_{\rm{ gen}}$, 
(\ref{eq:Generalized entropy at the t=0 slice}), 
as the function of $ x_0 $ ($ 0\leq x_0 \leq \pi/2 $) with 
$\phi_0 =1700,\, \beta=1,\, \Lambda=500,\, c=50 ,\, \varepsilon=0.1$. $\Delta=8\, (E=80)$(top) and 
$\Delta=1\, (E=10)$(bottom). The dotted line is the value of the entropy for the shock-less case, $\Delta=0\, (E=0)$. }}
\label{fig:plotgeneralizedat}
\end{figure}

By combining above results, we get the generalized entropy in the high temperature limit as the function of $x_0$, 
\begin{equation}
	\begin{aligned}
	S_{\text {gen }}\left(x_{0}\right)
	&=\left\{
	\begin{aligned}
		&\phi_{0}-\frac{(\pi X_\beta)^2}{\Lambda}  -2E\cos^2 x_0 \left( \frac{2 \pi X_\beta}{\Lambda}  +\frac{2}{\Lambda} E \cos ^{2} x_{0} -\tan x_{0}\right)+S_{L} &\text{for } x_0 < x_T\\[+10pt]
		&\phi_{0}-\frac{(\pi X_\beta)^2}{\Lambda}+S_{L}  \qquad\text{for }   x_T < x_0.
	\end{aligned}
	\right.
	\end{aligned}\label{eq:Generalized entropy at the t=0 slice}
\end{equation}
We plot the above generalized entropy for several values of $E$ in figure \ref{fig:plotgeneralizedat} \footnote{ The plots shown in this paper are obtained by full numerical calculations by faithfully extremizing the generalized entropies, on the contrary to analytical expressions appear in the body of the paper.}.

From the plot we see that when  the operator is  in the exterior of the horizon,  the  generalized entropy  does not change.  On the other hand, when it is inserted in the black hole interior  always make the entropy  decrease. Also we observe that, as the  location of the local operator goes deeper  interior of the black hole,  the entropy gets significantly decreased.  This is  because if the shock wave  is created inside of the horizon,   it makes the  interior wormhole region longer  (which is  seen from the relation $x_{H} (0) < x_{H} (E)$), and reduces the entropy of the black hole $\Phi_{\beta}( x_{H} (0) ) > \Phi_{\beta,E}( x_{H} (E) )$.    Therefore, in some sense what the shock wave does is to make the black hole further ``evaporate".   This black hole in the universe B has been evaporating due to the entanglement  with the non-gravitating universe A, and  the insertion of the local operator accelerates the evaporation, which leads to the faster decrease of the entropy.

\begin{figure}[htbp]
	\begin{minipage}{0.5\hsize}
    \vspace{0.3cm}
	\hspace{-2cm}
		\includegraphics[scale=0.6]{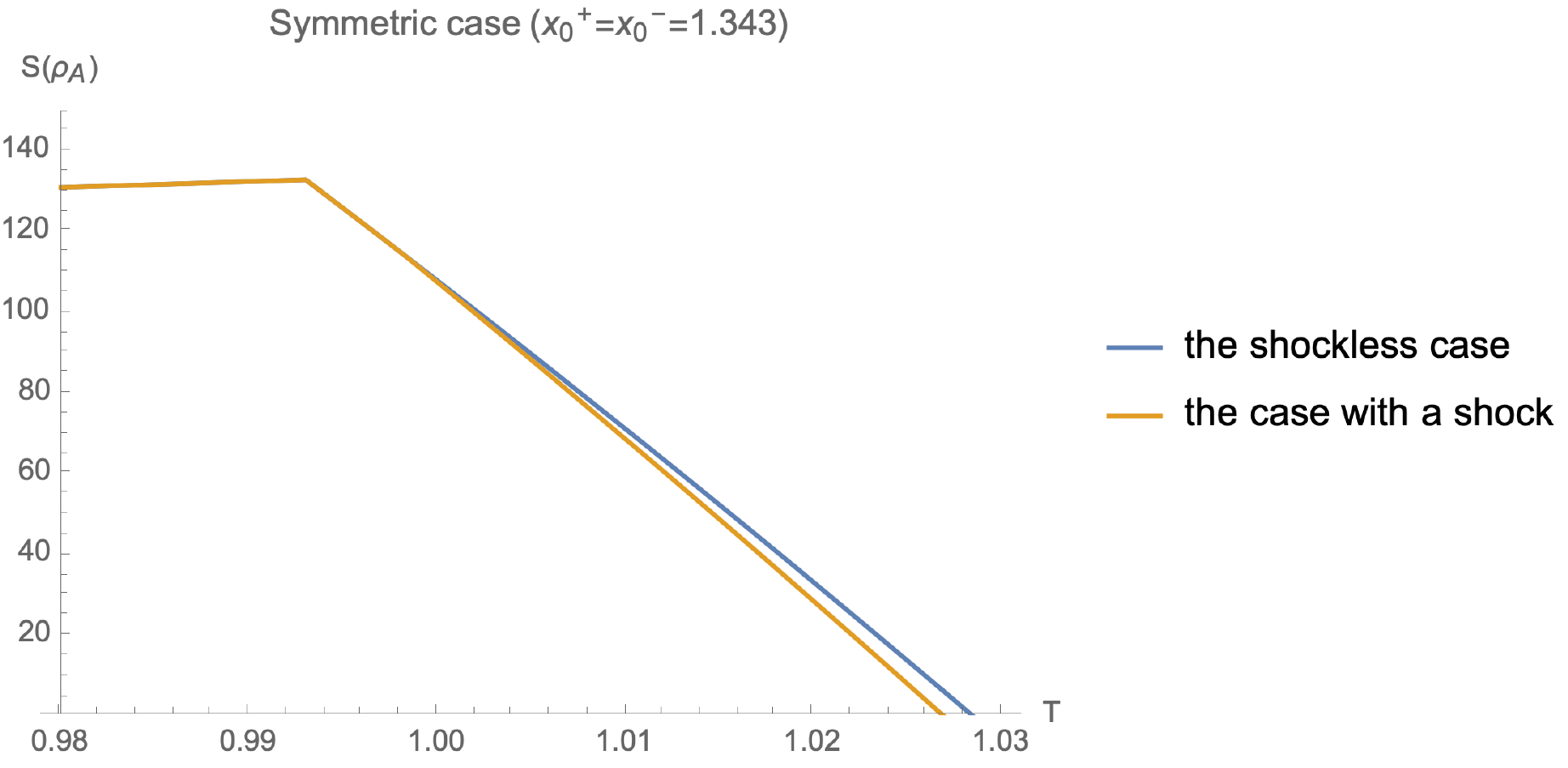}
    \end{minipage}
\centering
\caption{\small{Plots of the Page curves corresponding to the shockless case \eqref{eq:entwithoutshock} and the case with a shock  \eqref{eq:eeshockwave} as the function of $ T=1/\beta $ with fixing the position of the operator, which we place on the reflection symmetric slice $x_0^+=x_0^-$.   $ \phi_0 =1700,\, \Lambda=500,\, c=50,\, \Delta=10,\, \varepsilon=0.01,\, x_0^+=x_0^-=1.343$. The island begins dominating at $T\simeq 0.993$ and the location of the corresponding QES is $x_H^{+}(0)=x_H^{-}(0)\simeq 1.328$.  The entanglement entropy with a shock  decreases faster than the one without it.
}}
\label{fig:symmetricPage}
\end{figure}

The actual entanglement entropy $S(\rho_{A})$ is given by the minimum between $S_{{\rm no-island}}$ and $ S_{\text {gen }}\left(x_{0}\right)$. We plot this curve in figure \ref{fig:symmetricPage}. Since we are interested in how this $S(\rho_{A})$  changes as we increase the entanglement temperature $1/\beta$, we plot it as a function of  $1/\beta$ while the location of the operator $x^{\pm} =x^{\pm}_{0}$ is kept fixed. As we increase the entanglement temperature, the bifurcation surface of the black hole approaches the asymptotic infinity, and
the operator at $x^{\pm} =x^{\pm}_{0}$  is eventually absorbed into the black hole. We again observe that above some critical temperature, the entanglement entropy is dominated by the generalized entropy $S_{{\rm gen}} (x_{0})$ \eqref{eq:Generalized entropy at the t=0 slice}. In the high temperature limit, since  the operator goes into deep interior of the black hole, $S_{{\rm gen}} (x_{0})$  is given by the first line of  \eqref{eq:Generalized entropy at the t=0 slice}. Therefore an approximate expression for the  entanglement entropy reads,
\be
S(\rho_{A}) =
\begin{cases}
S_{{\rm no-island}} = \f{\pi^{2}c}{3\beta}  & \beta \gg \beta_{c}\\[+10pt]
  S_{{\rm gen}} (x_{0})= 
  	\phi_{0}-\frac{(\pi X_\beta)^2}{\Lambda}  -2E\cos^2 x_0 \left( \frac{2 \pi X_\beta}{\Lambda}  +\frac{2}{\Lambda} E \cos ^{2} x_{0} -\tan x_{0}\right)+S_{L}
   &\beta  \ll \beta_{c}.
\end{cases}
\label{eq:eeshockwave}
\ee

It is interesting to compare this result with the Page curve without the operator insertion \eqref{eq:entwithoutshock}.  In this case, above the critical temperature the entropy decreases as 
\be
S_{{\rm island}}=\phi_{0}-\frac{(\pi X_\beta)^2}{\Lambda}+S_{L}.
\ee

By comparing it with  \eqref{eq:eeshockwave}, we see that the entropy in the presence of the shock wave is reduced faster than the entropy without the shock, by increasing the entanglement temperature.  This also supports the point of view that the shock wave accelerates the evaporation of the black hole.

\subsection{QESs with non-trivial CFT entropy }
In the above examples, the CFT entropy did not play any role. This is because,  when the operator is inserted on the time reflection symmetric slice $x^{+}_{0} =x^{-}_{0}$, the  quantum extremal surfaces and the local operator are always spatially separated,  therefore the  non-trivial part of the CFT entropy  $\Delta S =S_{\beta, E}[\bar{C}] -S_{\beta}[\bar{C}]$ vanishes in the generalized entropy.   As a result, the  QESs coincide with the classical extremal surfaces, which can be identified with  the bifurcation surfaces of the black hole. 
On the other hand, when we insert the operator in the future light cone of the original horizon $x^{\pm}=x^{\pm}_{H}(0)$, then the non-trivial par of the CFT entropy is non vanishing. 
In this case, we insert the local operator in the region $x^{+}_{0}>x^{+}_{H}(0),  x^{-}_{0}<x^{-}_{H}(0)$. Let us first derive the location of the classical extremal surface. The dilaton profile  is still  given by \eqref{eq:completedl}, and since we expect the new bifurcation surface is  in the  past light cone of the operator,  we extremize 
\be
\begin{split}
\Phi_{R}(x^{\pm}) &= \phi_{0} + \f{\Lambda}{4} \tan x^{+}\tan x^{-} -X_{\beta} (x^{+} \tan x^{+} + x^{-} \tan x^{-} )\\
&-E \cos^{2} x^{-}_{0} (\tan x^{-}-\tan x^{-}_{0}).
\end{split}
\ee

The critical point  $(x^{+}_{\mathcal{H}} (E), x^{-}_{\mathcal{H}} (E))$  of this dilaton profile $\Phi_{R}(x^{\pm})$ satisfies
\be
x^{-}_{\mathcal{H}}(E)=x^{-}_{H}(0),\quad \tan x^{+}_{\mathcal{H}}(E)
=\tan x^{+}_{H}(0)  +\f{4E}{\Lambda} \cos^{2} x^{-}_{0} .
\ee
We remark that this is different from the critical point  $(x^{+}_{H} (E), x^{-}_{H} (E))$ of  $\Phi_{\beta, E}$  \eqref{eq:phibetaE} discussed in the previous subsection.

The net effect of the shock wave is shifting the horizon along the  $x^{+}$ direction. In order for this critical point $(x^{+}_{\mathcal{H}} (E), x^{-}_{\mathcal{H}} (E))$ to be really in the past light cone of the operator, we need, 
\be
\tan x^{+}_{H}(0)  +\f{4E}{\Lambda} \cos^{2} x^{-}_{0}< \tan x^{+}_{0}.
\ee

Now let us add quantum effects. The expression of the  generalized entropy can be obtained from \eqref{eq:futureentropy},
\be
S_{\rm{gen}} (x^{\pm}) =  \Phi_{R}(x^{\pm})+ \f{c}{6} \log \left[
\f{\beta}{\pi \ve} 
\f{\sin \pi \alpha}{\alpha}
\f{\sinh \f{\pi}{\beta} (x^{+}_{0}-x^{+}_{5})\sinh \f{\pi}{\beta} (x^{+}_{6}-x^{+}_{0})}
{\sinh \f{\pi}{\beta} (x^{+}_{6}-x^{+}_{5})}
\right]
+S_{\beta}[\bar{C}] -S_{\rm{vac}}[\bar{C}]+S_{L}
\label{eq:genentropy2}
\ee
where $S_{L}$ is the contribution of the left horizon, as in the previous subsection.  In the high temperature limit, $S_{L} =\Phi_{\beta} (x_{H}^{\pm}(0))$ defined in  \eqref{eq:dilatonvCES}.

We then specify the location of the quantum extremal surface $x^{\pm} =x^{\pm}_{Q_{1}}$ by finding the critical point of the  above generalized entropy \eqref{eq:genentropy2}. 
Since its derivative along the $x^{-}$ direction is not affected by the non-trivial part of the  CFT entropy $\Delta S$, 
$\tan x^{+}_{Q_{1}}$ is still given by 
\be
\tan x^{+}_{Q_{1}}
=\tan x^{+}_{H}(0)  +\f{4E}{\Lambda} \cos^{2} x^{-}_{0}.
\label{eq:newqesp}
\ee

The derivative along the $x^{+}$ direction  is modified by $\Delta S$. By ignoring its trivial part $S_{\beta}[\bar{C}] -S_{{\rm vac}}[\bar{C}]$, we get
\be
 \tan x_{Q_{1}}^{-}=\f{4}{\Lambda} \left[ \f{\pi}{2} X_{\beta} +\f{c\pi}{6\beta}\cos^{2} x_{Q_{1}}^{+}\left(\f{1}{\sinh \f{\pi}{\beta} (x_{0}^{+} -x_{Q_{1}}^{+})}- \f{1}{\sinh \f{\pi}{\beta} (x_{6}^{+} -x_{Q_{1}}^{+})}\right)\right].\label{eq:newqes}
\ee

The  contribution of this quantum extremal surface  is given by plugging the solution of these equations \eqref{eq:newqesp} and 
\eqref{eq:newqes}  to the expression of the  generalized entropy \eqref{eq:genentropy2}.

There is  another quantum extremal surface,  $x^{\pm} =x^{\pm}_{Q_{2}}$ located at the right wedge of the operator $x^{\pm}_{Q_{2}} > x^{\pm}_{0}$. In this case, the non-trivial part of the CFT entropy $\Delta S$ is vanishing, so it conincides with the bifurcation surface of the original black hole 
$x^{\pm}_{Q_{2}} =x^{\pm}_{H}(0)$.

Although we have two candidates of the quantum extremal surface at  $x^{\pm}=x^{\pm}_{Q_{1}}$ and $ x^{\pm}=x^{\pm}_{Q_{2}}$, they can not appear simultaneously. This is due to the non-symmetric insertion of the local operator. If we put the local operator at the future light cone of the bifurcation surface of the original black hole $x_0^{+}>x_{H}^{+}(0),\, x_0^{-}<x_{H}^- (0)$, then  the bifurcation surface is moved to $(x^{+}_{\mathcal{H}}(E), x^{-}_{\mathcal{H}}(E))$. In this case  $x^{\pm}_{H}(0)=x_{Q_2}^{\pm}$ is no longer extremal, and only  $x^{\pm} =x_{Q_1}^{\pm}$ is the quantum extremal surface. On the other hand, if the operator $x_{0}^{\pm}$  is in the exterior of the horizon $x_0^{\pm}>x_{H}^{\pm}(0)$,  then only $x^{\pm} =x_{Q_2}^{\pm}$ is the quantum extremal surface. Thereby, the generalized entropy is given by 
\be
S_{\rm{gen},E}(x_0^+,x_0^-)=
\begin{cases}
S_{\rm{gen}}(x^{\pm}_{Q_1}) \quad \text{for}\quad x_{H}^{-}(0) > x_0^{-}\\
S_{\rm{gen}}(x^{\pm}_{Q_2}) \quad \text{for}\quad x_{H}^{-}(0) < x_0^{-}.
\end{cases}
\label{eq:geneWishock}
\ee
In the high temperature limit, $S_{\rm{gen}}(x^{\pm}_{Q_1})$ is obtained by plugging the solution of \eqref{eq:newqes} into \eqref{eq:genentropy2}. $S_{\rm{gen}}(x^{\pm}_{Q_2}) $ coincides with the entropy of the original black hole $\Phi_{\beta}(x^{\pm}_{H}(0))$.
The actual entanglement entropy is given by the minimum between this generalized entropy and $S_{{\rm no-island}}$,
\be 
S(\rho_{A}) ={\rm min} \left\{ S_{{\rm no-island}}, \, S_{\rm{gen},E}\right\}.\label{eq:eewishoNonsyme}
\ee

\subsection*{Plot of the result}

\begin{figure}[t]
    \centering
    \includegraphics[width=7cm]{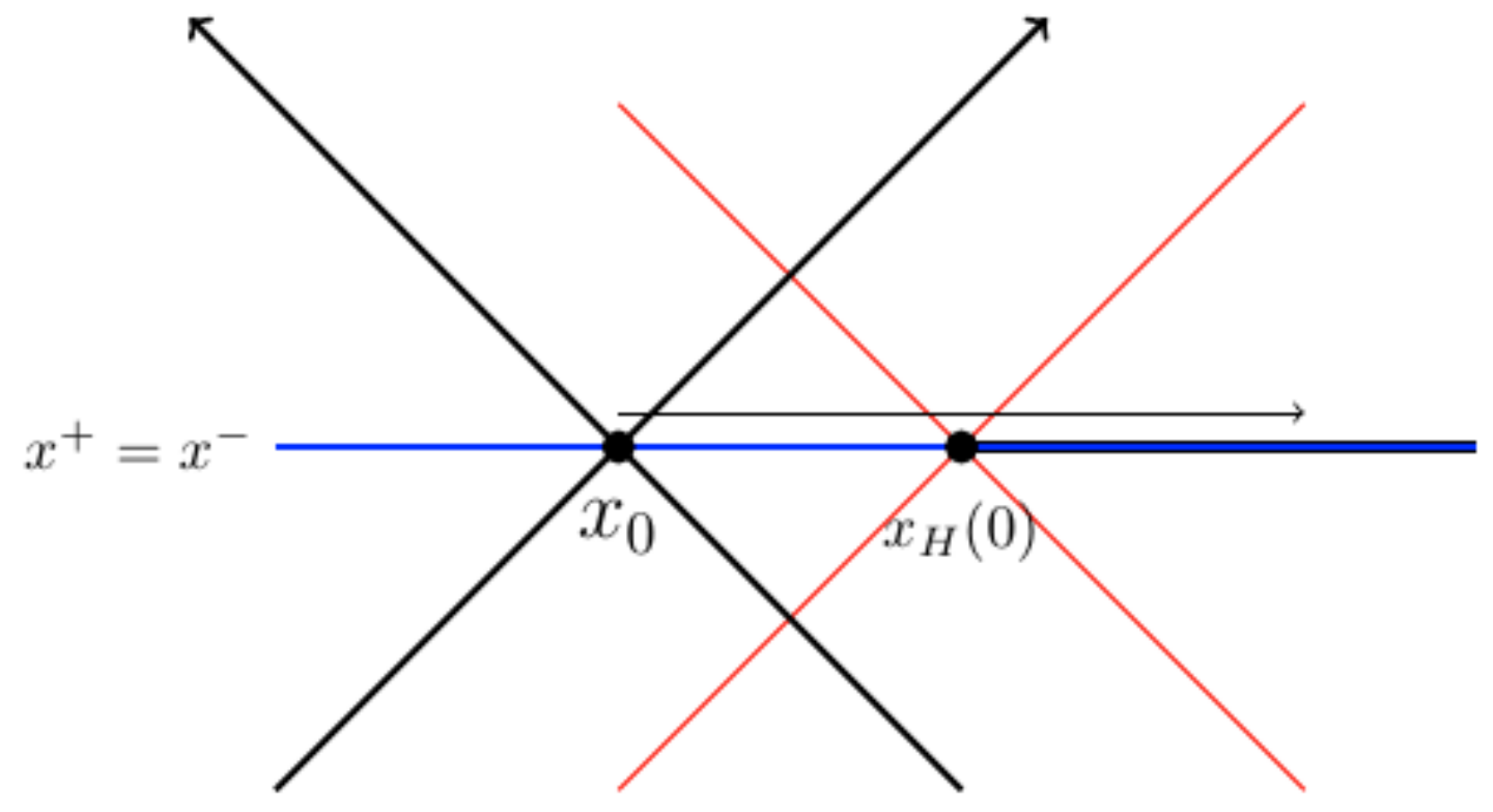}
    \hspace{0.5cm}
\includegraphics[width=6.5cm]{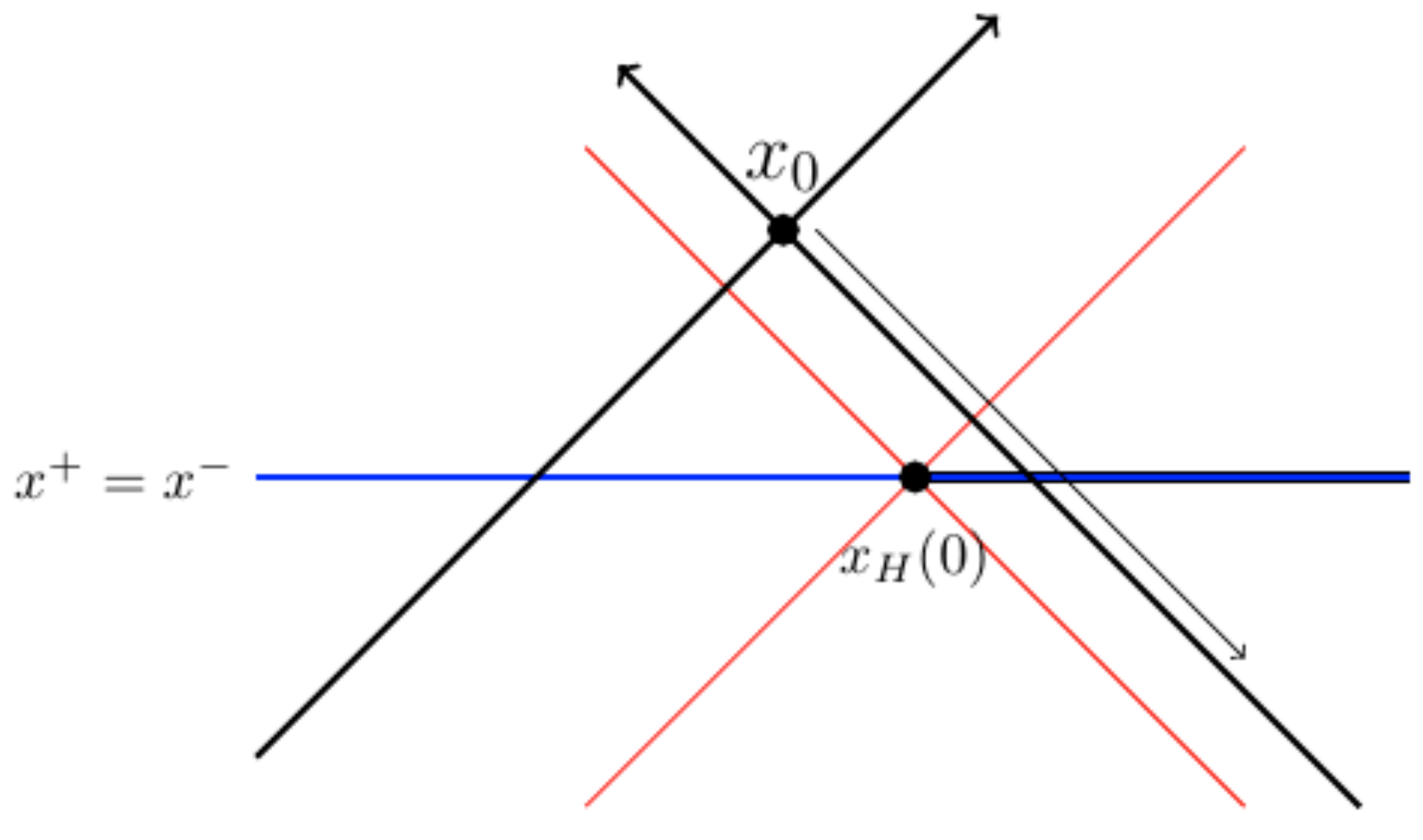}
  \hspace{0.5cm}
     \caption{\small{ {\bf Left}: The operator inserted on the time reflection symmetric slice (the blue line). In this case the operator is spatially separated from the bifurcation surface of the black hole at $x^{\pm} = x^{\pm}_{H} (0)$. (The black hole horizon is drawn by the red lines.) {\bf Right}: The operator inserted on the non time reflection symmetric slice. In this case, the bifurcation surface can causally contact with the operator. }}
    \label{fig:comparison}
\end{figure}

\begin{figure}[t]
    \centering
    \includegraphics[scale=0.8]{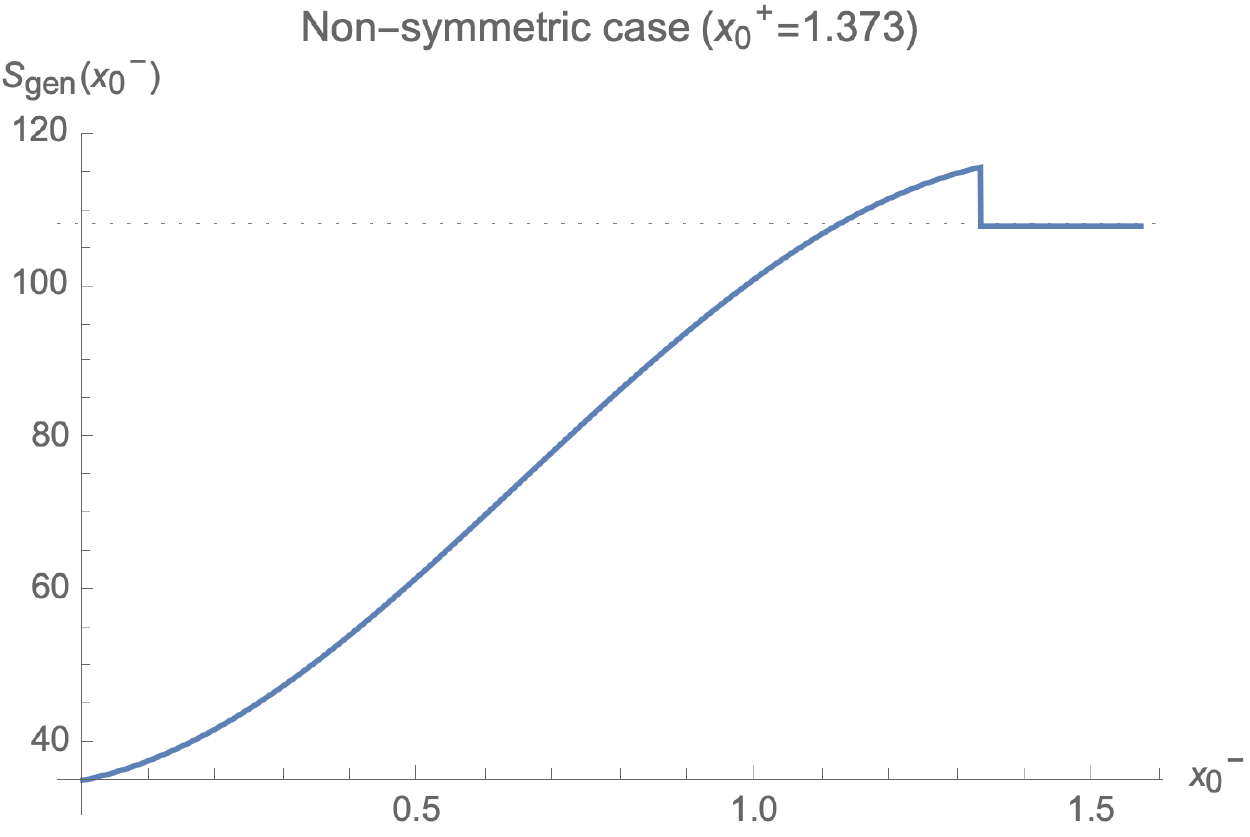}
    \hspace{0.5cm}
    \caption{\small{Plot of the generalized entropy given in \eqref{eq:geneWishock} as a function of $x_0^-$ with  $x_0^+$ kept fixed (the case of the non-symmetric insertion), with the choice of parameters $\phi_0=1700, \, \Lambda=500, \, c=50, \,\beta=1, \,\Delta=7, \,\varepsilon=0.1, \,x_0^+=1.373$. The dotted line corresponds to the shockless case $\Delta =0$.
}}
     \label{fig:nonsymmetric generalized entorpy}
\end{figure}

\begin{figure}[htbp]
	\begin{minipage}{0.5\hsize}
    \vspace{0.3cm}
	\hspace{-2cm}
		\includegraphics[scale=0.7]{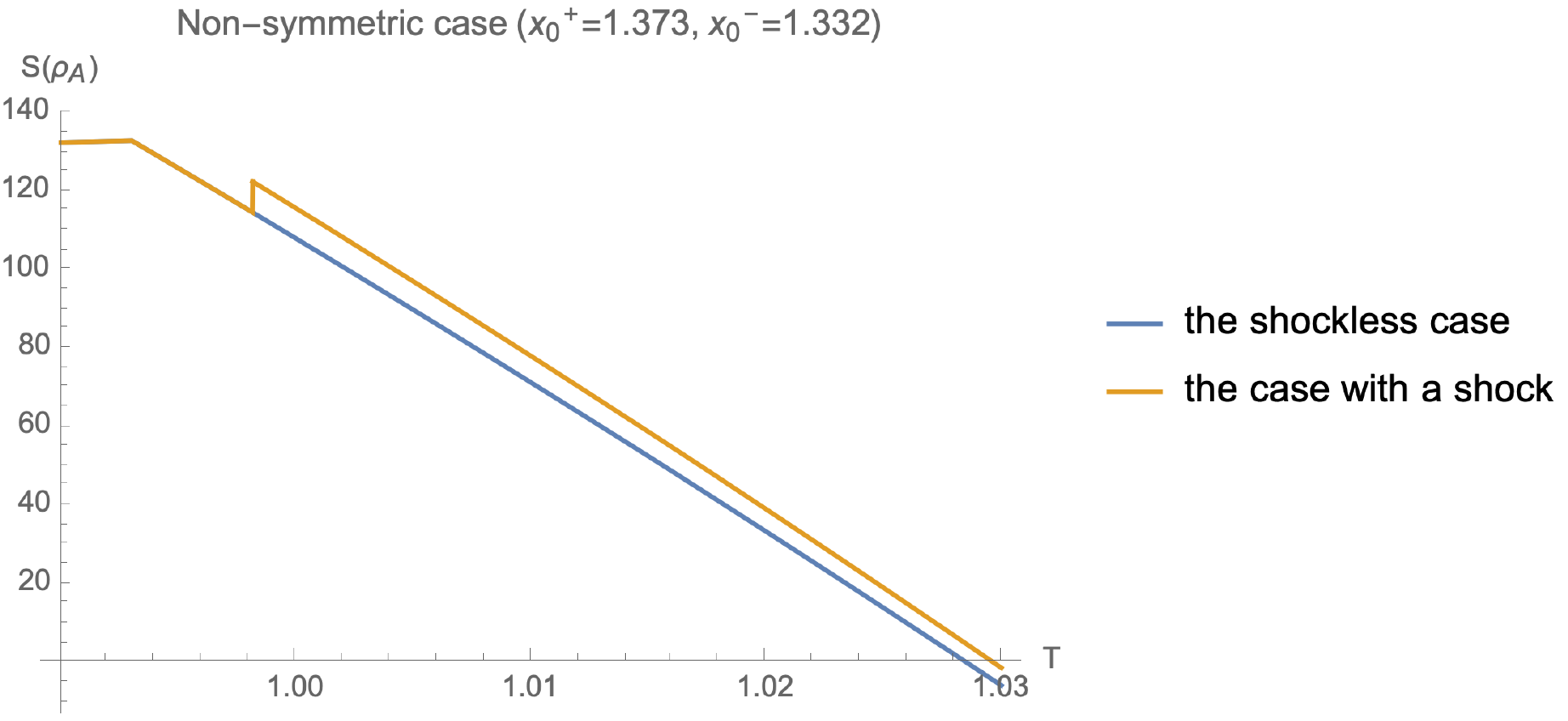}
    \end{minipage}
\centering
\caption{\small{Plots of the Page curves corresponding to the shockless case \eqref{eq:entwithoutshock} and the case with a shock wave \eqref{eq:eewishoNonsyme} as the function of $ T=1/\beta $ with fixing the position of the operator, which is not on the reflection symmetric slice, ie  $x_0^+\neq x_0^-$, with the choice of parameters   $ \phi_0 =1700,\, \Lambda=500,\, c=50,\, \Delta=7,\, \varepsilon=0.1,\, x_0^+=1.373,\, x_0^-=1.332$. The island  begins dominating at $T\simeq 0.993$ and the location of the corresponding QES is $x_H^{+}(0)=x_H^{-}(0)\simeq 1.328$. }}
	\label{fig:nonsmmetricPagecurve}
\end{figure}

\begin{figure}[t]
    \centering
    \includegraphics[scale=0.7]{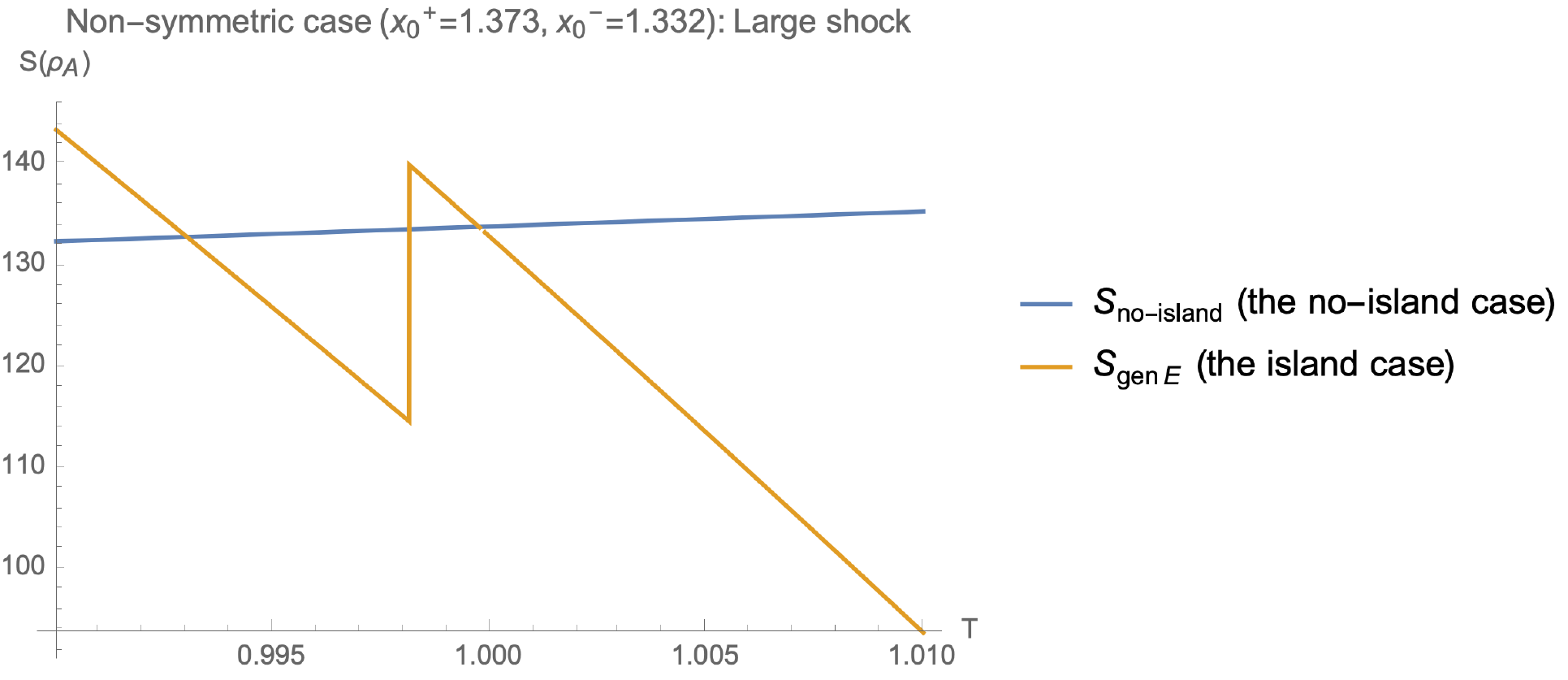}
    \hspace{0.5cm}
    \caption{\small{Similar plot to figure \ref{fig:nonsmmetricPagecurve} but the shock wave has a larger energy than the previous case. We plot the Page curve only around the points at which the non-trivial dominance changes happen unlike the previous case (figure \ref{fig:nonsmmetricPagecurve}). The Page curve is given by the minimum of them. We set the parameters to  $ \phi_0 =1700,\, \Lambda=500,\, c=50,\, \Delta=8,\, \varepsilon=0.01,\, x_0^+=1.373,\, x_0^-=1.332$. In this case, the  transitions between them happen several times. The first transition is at $T\simeq 0.993$ and the location of the corresponding QES is $x_H^{+}(0)=x_H^{-}(0)\simeq 1.328$. }}
    \label{fig:ThePagecurvemultiple}
\end{figure}

Let us focus on the case, where 
 the location of the left mover  $x^{+} =x^{+}_{0}$ is fixed, but   the location of the right mover   $x^{-} =x^{-}_{0}$ is varied as in the right panel of figure  \ref{fig:comparison}.  We also  demand $x^{+}_{0} > x^{+}_{H}(E)$, so that the location of the operator interpolates the  interior and the exterior of the black hole.  We  plot  the generalized entropy $S_{{\rm gen},E} (x^{\pm}_{0})$  in figure \ref{fig:nonsymmetric generalized entorpy}.
 By decreasing the value of $x^{-}_{0}$,  the local operator is falling to the black hole horizon, and  we are interested in, how the entropy changes as the local operator is falling to the horizon and eventually  enters the interior\footnote{
 In studying  the Page curve for an evaporating  black hole,  the setup where  an AdS black hole is attached to a non-gravitating  heat bath at the asymptotic boundary is often used, for example in \cite{Almheiri:2019psf,Hollowood:2020cou,Goto:2020wnk}.  In such a setup, the local operator itself is inserted in the (non-gravitating) bath region, and the shock wave created by the operator can enter the bulk region. Instead, in our setup,
 we  insert the operator in the gravitating universe, and the operator itself can enter the black hole interior.  We regard our shock waves  is a kind of  `heavy diaries', ex.\cite{Penington:2019npb,Chen:2019uhq,Hollowood:2020cou}. }.

 This plot  in figure \ref{fig:nonsymmetric generalized entorpy} for the asymmetric insertion  is compared to the similar plot shown in figure \ref{fig:plotgeneralizedat}, where we insert the operator on the time reflection symmetric slice \eqref{eq:Generalized entropy at the t=0 slice}, as in the left panel of  figure  \ref{fig:comparison}.
These two plots share a common feature. Namely, when the operator is inserted outside of the horizon $x^{-}>x^{-}_{H}(0)$,  the values of these two entropies both approach the classical  entropy  of the original black hole $\Phi_{\beta} (x^{\pm}_{H}(0))$, defined in \eqref{eq:dilatonvCES}. This happens because, the non-trivial part of the CFT entropy $\Delta S$ vanishes  since the local operator is in the causal diamond of $\bar{C}$.

However these two generalized entropies behave differently when the operator is inserted in the black hole interior $x^{-}_{0}<x^{-}_{H}(0) $.  In particular, there is a bump in  the entropy plot  for the asymmetric insertion in figure \ref{fig:nonsymmetric generalized entorpy}, which  is absent in the plot for the  symmetric insertion \ref{fig:plotgeneralizedat}. This difference is a direct consequence of  the fact that the quantum extremal surface for the asymmetric insertion is in the past of the local operator, so the non-trivial  part of the  CFT entropy  $\Delta S$ is non vanishing.  On the other hand, in the case of the  symmetric insertion, the QES is spatially separated from the local operator, so 
 $\Delta S$ is vanishing.

This bump in figure \ref{fig:nonsymmetric generalized entorpy}  can be understood as a result of the dynamics of the black hole.
For this purpose, it is useful to follow the plot backward in the $x^{-}_{0}$ direction.
By decreasing $x^{-}_{0}$, the black hole gets bigger due to its absorption of  the local operator, and this causes the sudden increase of the entropy in the plot. After the increase of its size, the black hole starts to evaporate again, and as a result, the generalized  entropy starts decreasing.

One can  characterize the difference between these two generalized
entropies, by the difference of the natures of these two insertions. The symmetric insertion can be regarded as a local operation, because in this case the local operator is always in either the causal diamond of the island $D[C]$ (which can be regarded as a part of the radiation system) or the entanglement wedge of the black hole $D[\bar{C}]$, as in the left panel of figure \ref{fig:comparison}. Since they are LOCCs,  they can only decrease the entanglement entropy. On the other hand,  operators inserted asymmetrically  enter a region of the black hole interior, which does not belong to neither 
of these two entanglement wedges (the right panel of figure  \ref{fig:comparison}). Therefore, these insertions are not LOCCs, so they can increase the entanglement between  two wedges.

\subsection{The entanglement entropy}

The entanglement entropy $S(\rho_{A})$ is given by putting these results to the formula \eqref{eq:eewishoNonsyme}. 
Again we plot this as a function of the entanglement temperature $1/\beta$ with the location of the operator $x^{\pm}=x^{\pm}_{0}$ kept fixed  in figure \ref{fig:nonsmmetricPagecurve} and  \ref{fig:ThePagecurvemultiple}.  As we increase temperature, the horizon expands, and the local operator is absorbed into the black hole. So also in this case, we can see the identical physics which leads to the result obtained by varying $x^{-}_{0}$.     

When the location of the operator is properly chosen, 
the resulting entanglement entropy behaves in a complicated manner as in \ref{fig:ThePagecurvemultiple}. This is  compared to the same entropy without the shock wave \eqref{eq:entwithoutshock}, where the transition between    $S_{{\rm no-island}}$  and $S_{{\rm island}}$ happens only once.
Instead, in the presence of the shock, the transition can happen multiple times. In the  actual plot in figure \ref{fig:ThePagecurvemultiple}, we  observe that at sufficiently low temperature,  $S_{{\rm no-island}}$ dominates, and by increasing the entanglement temperature  $S_{{\rm island}}$ becomes the dominant one, as we can also see in the case without the shock wave. However, this is not the end of the story. Namely further increase of the temperature  makes the horizon expand, so the local operator is falling to the horizon. This will lead to the size change of the black hole, and to the sudden increase of  $S_{{\rm island}}$.  Now this  $S_{{\rm island}}$ gets larger than the naive Hawking's entropy, so above this temperature $S_{{\rm no-island}}$ again dominates. After this,  the size of the black hole is eventually reduced due to the emissions of Hakwing quanta, so eventually 
$S_{{\rm island}}$ gets smaller than  $S_{{\rm no-island}}$, and it becomes dominant once  again.

\section{Conclusion}
\label{section:conc}

 In this paper, we studied  dynamics of  black holes in flat space, when it is entangled with an auxiliary non-gravitating universe. We find the back-reaction of the entanglement between them reduces  the horizon area of the black hole, and lengthening  its interior region. This lengthening  can be understood in terms of   monogamy property of  entanglement \cite{Balasubramanian:2020coy}.  Since the gravitating universe B contains two horizons,  the Hilbert space of $B$  can be naturally  decomposed into two  horizon Hilbert spaces $H_{B_{L}} \otimes H_{B_{R}}$. Since both of these degrees of freedom are strongly entangled with $H_{A}$, the entanglement  between two horizons should be suppressed, according to  monogamy of entanglement. This suppression is geometrized by the long interior  region of the black hole.  We then computed the entanglement entropy between the two universes, and found  a Page curve for an evaporating black hole. 

 We also studied  actions of local operations on the black hole. Such a local operation is modeled by an insertion of a CFT operator.  In our setup, it is natural to consider the insertions in the black hole interior, in addition to the ones in the exterior. This insertion can back-react to the black hole through its stress energy tensor.  The (quantum) extremal surfaces  in the back-reacted black hole highly depend on the location of the insertion.   There  are several differences between the insertions in the interior and  exterior. When the operator is in the exterior of the black hole, it does not change the entanglement entropy. On the other hand, the entropy is significantly reduced when the  operator is in the interior. The disruption becomes stronger as the location of  the insertion  gets deeper in the interior of the black hole.

 It would be interesting to study how can  an observer in the non-gravitating universe A   recover the information of shock wave  in the black hole interior  in the gravitating universe B. One way to do so  is using the modular flow   of the reduced density matrix  $\rho_{A}$ .  Our setup is especially  suitable for this purpose.  This is because  these two universes can be embedded in the larger Minkowski space, in which AB are both realized as the left and right wedges of the origin.  Furthermore, the TFD state on $AB$ is identified with  the vacuum of this larger Minkowski space. By focus on the code subspace, the modular Hamiltonian  of $\rho_{A}$  is approximated by the CFT vacuum modular Hamiltonian of the CFT on $AC$, $-\log \rho_{A} = K^{{\rm CFT}}_{AC}$, where $C$ is the island region in the gravitating universe B \cite{Chen:2019iro}. In a CFT with a large central charge and a sparse spectrum, this modular Hamiltonian  is given by the sum of two modular Hamiltonians for single intervals, each of which connects the endpoints of A and C. In 2d CFT, this vacuum modular Hamiltonian has a particularly simple form, as an integral of stress energy tensor. Since this  modular flow is geometric, one can visualise how an operator in the black hole interior 
 gets out of the horizon under the flow.

\section*{Acknowledgement}

TU  thanks Vijay Balasubramanian, Arjun Kar and Kotaro Tamaoka for useful discussions in the related projects. TU was supported by JSPS Grant-in-Aid for Young Scientists  19K14716.  

\appendix

\section{Entanglement entropy and local quench for two disjoint intervals } \label{section:EEderivation}

In this  appendix we derive the results presented in  \ref{subsubsection:CFTentropy} and \ref{subsubsection:CFTdisjoint} for the CFT entanglement entropy in the presence of a local operator insertion, by following the argument of \cite{Caputa:2015waa}. We are interested in the state \eqref{eq:exstate} of the total system $AB$, and its reduced density matrix on two disjoint intervals $\bar{C}$ in the universe $B$. In the body of this paper, this region $\bar{C}$ is identified with the compliment of the island $C$ in the universe B.

\begin{figure}[t]
	\centering
	\includegraphics[width=10cm]{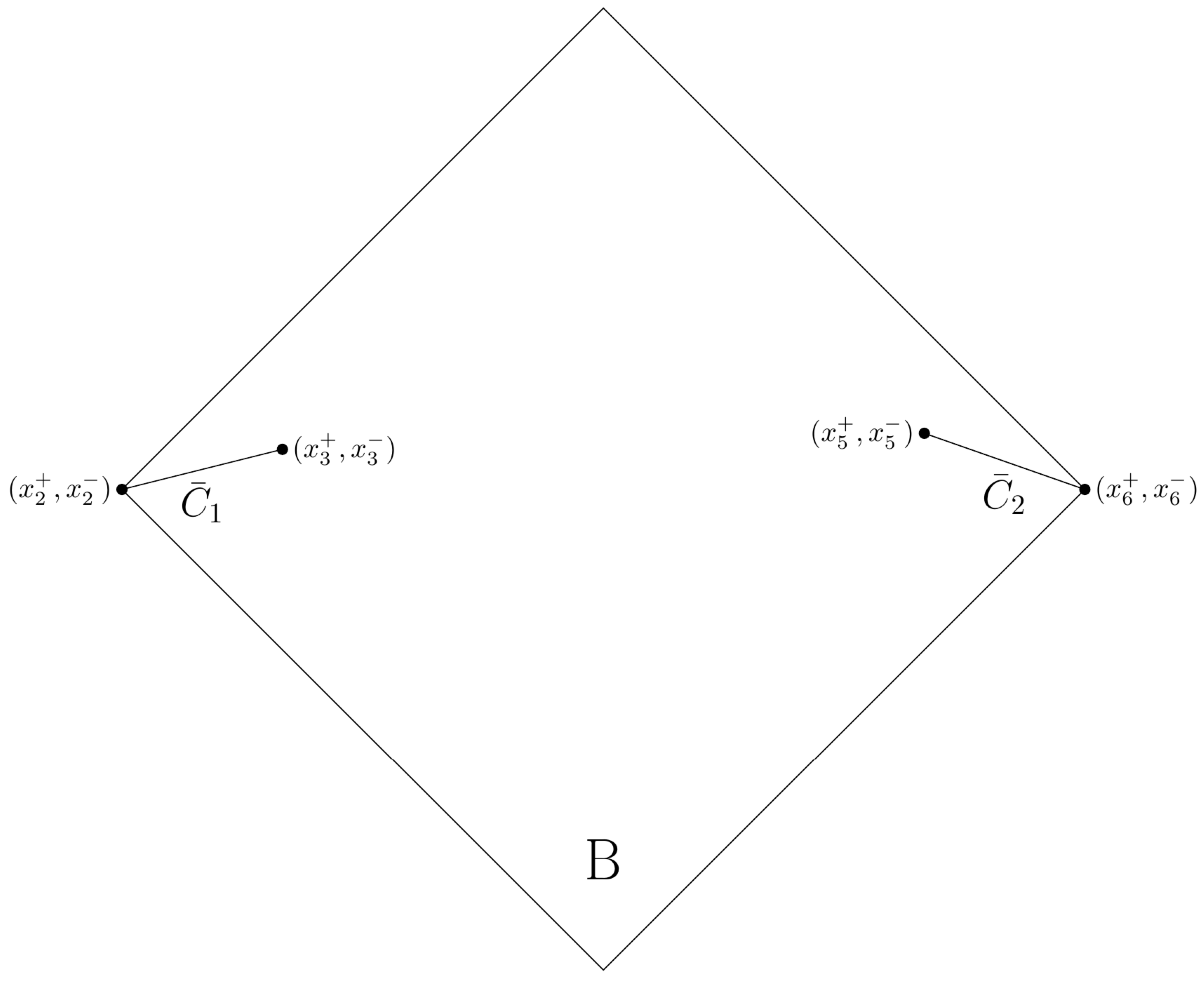}
	\hspace{0.5cm}
	
	\hspace{0.5cm}
	\caption{\small{The two intervals in the universe $B$}}
	\label{fig:setupCbar}
\end{figure}

Let $ \rho_{\bar{C}}$ denote the reduced density matrix on two disjoint intervals ${\bar{C}}={\bar{C}}_1\cup {\bar{C}}_2$, whose endpoints are given by $ x_2^\pm$ and $ x_3^\pm$ ($ x_2^\pm\leq x_3^\pm $) for $ {\bar{C}}_1 $ and $ x_5^\pm$ and $ x_6^\pm$ ($x_3^\pm < x_5^\pm\leq x_6^\pm $) for $ {\bar{C}}_2 $ respectively(see figure \ref{fig:setupCbar}), that is, $ \rho_{\bar{C}} = \mathrm{tr}_{{C}}\; \rho$, where $\rho$ is given by the reduced density matrix of the  universe B \eqref{eq:denmat}.
The CFT entanglement entropy of the density matrix can be calculated by using the replica trick. For this purpose, we first consider the $n$-th R\'{e}nyi  entanglement entropy
\begin{equation}
	S^{(n)}[\bar{C}] \equiv \frac{1}{1-n} \log \operatorname{tr} \rho_{\bar{C}}^{n}
\end{equation}
and, by taking the limit 
\begin{equation}
	S[\bar{C}]=\lim_{n\to 1}  	S^{(n)}[\bar{C}] ,
\end{equation}
we obtain the CFT entanglement entropy.

To evaluate   $ \operatorname{tr} \rho_{\bar{C}}^{n} $, we need to compute a normalized $ 2n $-point function on an $ n $-sheeted replica manifold branched along the $ \bar{C} $. Since the reduced density matrix $\rho_{\bar{C}}$ has the  thermal form, each replica sheet is a cylinder with the period  $\beta$.   This $2n$-point function is identical to the  normalized six-point function including twist operators on a (no-replicated) manifold (thermal cylinder), in the cyclic orbifold theory $CFT^{\otimes n}/Z_{n}$. We adopt the later description  and compute the six-point function
\begin{equation}
	\operatorname{tr} \rho_{\bar{C}}^{n} =\frac{\langle \mathcal{O}^{\otimes n}(x_1^+,x_1^-)\sigma_n(x_2^+,x_2^-)\sigma_{-n}(x_3^+,x_3^-)\sigma_n(x_5^+,x_5^-)\sigma_{-n}(x_6^+,x_6^-) \mathcal{O}^{\dagger \otimes n}(x_4^+,x_4^-)\rangle_{\beta}}{(\langle \mathcal{O}(x_1^+,x_1^-)\mathcal{O}^\dagger(x_4^+,x_4^-) \rangle_{\beta})^n},\label{eq: twist field correlation function}
\end{equation}
where $\langle \cdots \rangle_{\beta}$ means the thermal trace ${\rm tr} [\;\rho_\beta\  \cdots]$.

In the above correlation function,  we introduced the UV regulator $ \varepsilon $  in the location of the operators as follows,
\begin{equation}
	\left\{
	\begin{aligned}
		x_{1}^{\pm}&=x_{0}^{\pm} \mp i \varepsilon, \\
		x_{4}^{\pm}&=x_{0}^{\pm} \pm i \varepsilon;
	\end{aligned}
	\right.\label{eq:x1x4andx0}
\end{equation}
$  \mathcal{O}^{\otimes n} $ and $ \mathcal{O}^{\dagger \otimes n} $ represent the products of the operators $ \mathcal{O}_i $ and $ \mathcal{O}^\dagger_i $ which are the $i$-th copies of the operators in the cyclic orbifold theory,
\begin{equation}
	\begin{aligned}
		\mathcal{O}^{\otimes n}&=\mathcal{O}_{1} \mathcal{O}_{2} \ldots \mathcal{O}_{n},\\
		\mathcal{O}^{\dagger \otimes n}&=\mathcal{O}_{1}^{\dagger} \mathcal{O}_{2}^{\dagger} \ldots \mathcal{O}_{n}^{\dagger}.
	\end{aligned}
\end{equation}
 with the conformal dimension $ nh_\mathcal{O} $; $ \sigma_n $ and $ \sigma_{-n} $ are twist and anti-twist fields respectively with the conformal dimension $2H_n$,
\begin{equation}
	H_n=\frac{c}{24}\left(n-\frac{1}{n} \right) .
\end{equation}

We will compute the six-point function \eqref{eq: twist field correlation function} in  a conformal field theory with gravity dual. Such a CFT has a  large central charge and a sparse spectrum. In this class of theories, one can approximate the correlation function by a six-point Virasoro vacuum conformal block with  an appropriate choice of branch,  following the argument of \cite{Caputa:2015waa}. In doing so, it is convenient to map the thermal cylinder  to a plane by 
\begin{equation}
	w^{\pm}(x^\pm)=\exp\left(\frac{2 \pi}{\beta}\left(x^{\pm}-x_{0}^{\pm}\right)\right)
\end{equation}
and additionally apply a conformal transformation
\begin{equation}
	z^{\pm}(w^{\pm})=\frac{\left(w_{1}^{\pm}-w^{\pm}\right) w_{34}^{\pm}}{w_{13}^{\pm}\left(w^{\pm}-w_{4}^{\pm}\right)},
\end{equation}
where we use the notation $w_{i j}^{\pm}=w_{i}^{\pm}-w_{j}^{\pm}$ .
By the conformal map $ w^\pm \to z^\pm $, 
one can relate  $\operatorname{tr} \rho_{\bar{C}}^{n}$ in (\ref{eq: twist field correlation function})  to the  correlator on the  plane, 
\begin{equation}
	\begin{split}
		\operatorname{tr} \rho_{\bar{C}}^{n}=&((1-z^+)(1-z^-))^{2H_n}(z_{65}^+z_{65}^-)^{2H_n}\\
		&\quad\times\left\lbrace\left(\frac{\beta}{\pi\varepsilon_{UV}} \right)^4 \sinh\left(  \frac{\pi}{\beta} x_{65}^{+} \right) \sinh\left(  \frac{\pi}{\beta} x_{65}^{-}\right) \sinh\left(  \frac{\pi}{\beta} x_{32}^{+} \right) \sinh\left(  \frac{\pi}{\beta} x_{32}^{-}\right)   \right\rbrace ^{-2H_n}\\
		&\qquad\times \langle \mathcal{O}^{\otimes n}\left|\sigma_{n}\left(z^+, z^{-}\right) \sigma_{-n}(1,1) \sigma_{n}\left(z_{5}^{+}, z_{5}^{-}\right)\sigma_{-n}\left(z_{6}^{+}, z_{6}^{-}\right)\right| \mathcal{O}^{\otimes n}\rangle_,
	\end{split}\label{eq: conformal-tranformed twist field correlation function}
\end{equation}
where we introduce the UV cutoff $ \varepsilon_{UV} $ and the notation
\begin{equation}
	\begin{split}
		&\langle \mathcal{O}^{\otimes n}\left|\sigma_{n}\left(z^+, z^{-}\right) \sigma_{-n}(1,1) \sigma_{n}\left(z_{5}^{+}, z_{5}^{-}\right)\sigma_{-n}\left(z_{6}^{+}, z_{6}^{-}\right)\right| \mathcal{O}^{\otimes n}\rangle\\
		&\equiv\lim _{z^+_{4},z^-_{4} \rightarrow \infty}\left(z^+_{4}z^-_{4}\right)^{2 nh_{\mathcal{O}}}\langle  \mathcal{O}^{\dagger \otimes n}(z_4^+,z_4^-) \sigma_n(z^+,z^-)\sigma_{-n}(1,1)\sigma_n(z_5^+,z_5^-)\sigma_{-n}(z_6^+,z_6^-) \mathcal{O}^{\otimes n}(0,0)\rangle. \label{eq: 6-pt. correlation function}
	\end{split}
\end{equation}

Next, we evaluate (\ref{eq: 6-pt. correlation function}) with an insertion of a complete set as follows
\begin{equation}
	\begin{split}
		&\langle \mathcal{O}^{\otimes n}\left|\sigma_{n}\left(z^+, z^{-}\right) \sigma_{-n}(1,1) \sigma_{n}\left(z_{5}^{+}, z_{5}^{-}\right)\sigma_{-n}\left(z_{6}^{+}, z_{6}^{-}\right)\right| \mathcal{O}^{\otimes n}\rangle\\
		&=\sum_{\alpha}\langle \mathcal{O}^{\otimes n}\left|\sigma_{n}\left(z^+, z^{-}\right) \sigma_{-n}(1,1)|\alpha\rangle \langle\alpha| \sigma_{n}\left(z_{5}^{+}, z_{5}^{-}\right)\sigma_{-n}\left(z_{6}^{+}, z_{6}^{-}\right)\right| \mathcal{O}^{\otimes n}\rangle,
	\end{split}\label{eq:  6-pt. correlation function with a complete set}
\end{equation}
where the sum runs over all possible intermediate states.
However, in the $ \varepsilon\to 0 $ limit,  $ z^\pm $ approach $ 1 $,  and since the OPE  $\sigma_{n} (z) \sigma_{-n}(1)$ starts from the identity,  one can approximate the six-point function as a product of four-point functions, 
\begin{equation}
	\begin{split}
		&\langle \mathcal{O}^{\otimes n}\left|\sigma_{n}\left(z^+, z^{-}\right) \sigma_{-n}(1,1) \sigma_{n}\left(z_{5}^{+}, z_{5}^{-}\right)\sigma_{-n}\left(z_{6}^{+}, z_{6}^{-}\right)\right| \mathcal{O}^{\otimes n}\rangle\\
		&\simeq \langle \mathcal{O}^{\otimes n}\left|\sigma_{n}\left(z^+, z^{-}\right) \sigma_{-n}(1,1)|\mathcal{O}^{\otimes n}\rangle \langle\mathcal{O}^{\otimes n}| \sigma_{n}\left(z_{5}^{+}, z_{5}^{-}\right)\sigma_{-n}\left(z_{6}^{+}, z_{6}^{-}\right)\right| \mathcal{O}^{\otimes n}\rangle,
	\end{split}\label{eq: approximated 6-pt. function}
\end{equation}
in the $\ve \rightarrow 0$ limit.
By further applying a  conformal map
\begin{equation}
	\tilde{z}^{\pm}\left(z^{\pm}\right)=\frac{\left(z_{1}^{\pm}-z^{\pm}\right) z_{64}^{\pm}}{z_{16}^{\pm}\left(z^{\pm}-z_{4}^{\pm}\right)}
\end{equation}
to the second four-point function in \eqref{eq: approximated 6-pt. function}, we get
\begin{equation}
	\begin{split}
		&\operatorname{tr} \rho_{\bar{C}}^{n}(t)\\
		&=\quad\left\lbrace\left(\frac{\beta}{\pi\varepsilon_{UV}} \right)^4 \sinh\left(  \frac{\pi}{\beta} x_{65}^{+} \right) \sinh\left(  \frac{\pi}{\beta} x_{65}^{-}\right) \sinh\left(  \frac{\pi}{\beta} x_{32}^{+} \right) \sinh\left(  \frac{\pi}{\beta} x_{32}^{-}\right)   \right\rbrace ^{-2H_n}\\
		&\qquad\qquad\times((1-z^+)(1-z^-))^{2H_n} \left\langle\mathcal{O}^{\otimes n}\left|\sigma_{n}\left(z^{+}, z^{-}\right) \sigma_{-n}(1,1)\right| \mathcal{O}^{\otimes n}\right\rangle\\
		&\qquad\qquad\quad\times\left(\left(1-\tilde{z}_{5}^{+}\right)\left(1-\tilde{z}_{5}^{-}\right)\right)^{2 H_{n}}\left\langle\mathcal{O}^{\otimes n}\left|\sigma_{n}\left(\tilde{z}_{5}^{+}, \tilde{z}_{5}^{-}\right) \sigma_{-n}(1,1)\right| \mathcal{O}^{\otimes n}\right\rangle.
	\end{split}\label{eq: decoupled conformal-tranformed twist field correlation function}
\end{equation}

Generally, it is difficult to get a complete analytic expression of the four-point functions since they depend on the details of the dynamics of the theory.
However, because we focus on the theory that has the large central charge $ c \gg 1 $ and the sparse spectrum, the four-point functions can be well approximated by the vacuum Virasoro conformal  blocks. Moreover, in evaluating the entanglement entropy by taking $n\rightarrow 1$ limit of the twist operators in the correlator \eqref{eq: decoupled conformal-tranformed twist field correlation function}, we only need the Heavy-Heavy-Light-Light  Virasoro blocks, because  the conformal dimension of the twist operators becomes light, $H_{n} \rightarrow 1, n\rightarrow 1$.
Upon taking the limit, we keep the conformal dimension $ h_\mathcal{O} $ of the local operator $\mathcal{O}$ which we assume to be proportional to the central charge $c$ fixed. 
The dominant contribution of such a four-point function under the limit is given by \cite{Fitzpatrick:2014vua}
\begin{equation}
	\left(\left(1-z^{+}\right)\left(1-z^{-}\right)\right)^{2 H_{n}}\left\langle\mathcal{O}^{\otimes n}\left|\sigma_{n}\left(z^{+}, z^{-}\right) \sigma_{-n}(1,1)\right| \mathcal{O}^{\otimes n}\right\rangle \simeq\left(\frac{(z^+)^{\frac{1-\alpha}{2}}\left(1-(z^+)^{\alpha}\right) (z^-)^{\frac{1-\alpha}{2}}\left(1-(z^-)^{\alpha}\right)}{\alpha^{2}(1-z^+)(1-z^-)}\right)^{-2 H_n},
\end{equation}
where $\alpha=\sqrt{1-\frac{24 h_{\mathcal{O}}}{c}} =\sqrt{1-\frac{12 \Delta}{c}}$.
By plugging this result in \eqref{eq: decoupled conformal-tranformed twist field correlation function}  and taking $n\rightarrow 1$ limit, we get, the result for the CFT entanglement entropy $S[\bar{C}] =S_\beta[\bar{C}] + \Delta S [\bar{C}]$ with

\begin{equation}
	\begin{aligned}
		S_\beta[\bar{C}]&= \frac{c}{6} \log \left[\left(\frac{\beta}{\pi \varepsilon_{U V}}\right)^{4} \sinh \left(\frac{\pi}{\beta} x_{65}^{+}\right) \sinh \left(\frac{\pi}{\beta} x_{65}^{-}\right) \sinh \left(\frac{\pi}{\beta} x_{32}^{+}\right) \sinh \left(\frac{\pi}{\beta} x_{32}^{-}\right)\right]\\
		\Delta S [\bar{C}]&=\frac{c}{6} \log \left[\frac{\left(z^{+}\right)^{\frac{1-\alpha}{2}}\left(1-\left(z^{+}\right)^{\alpha}\right)\left(z^{-}\right)^{\frac{1-\alpha}{2}}\left(1-\left(z^{-}\right)^{\alpha}\right)}{\alpha^{2}\left(1-z^{+}\right)\left(1-z^{-}\right)}\right] \\
		&\quad+\frac{c}{6} \log \left[\frac{\left(\tilde{z}_{5}^{+}\right)^{\frac{1-\alpha}{2}}\left(1-\left(\tilde{z}_{5}^{+}\right)^{\alpha}\right)\left(\tilde{z}_{5}^{-}\right)^{\frac{1-\alpha}{2}}\left(1-\left(\tilde{z}_{5}^{-}\right)^{\alpha}\right)}{\alpha^{2}\left(1-\tilde{z}_{5}^{+}\right)\left(1-\tilde{z}_{5}^{-}\right)}\right].
	\end{aligned}\label{eq:ordinary entanglement entropy + shock part}
\end{equation}
$ S_\beta[\bar{C}] $ is the CFT entanglement entropy of the two disjoint intervals at finite temperature $ T=1/\beta $ and $\Delta S [\bar{C}] $ is the contribution to the CFT entanglement entropy from the perturbation by the local operator $ \mathcal{O} $ \footnote{Note that if we choose the other channel, then $ S_\beta[\bar{C}] $ and $\Delta S [\bar{C}] $ take a different form.}.

The right hand side of  the above formula  \eqref{eq:ordinary entanglement entropy + shock part} contains branch cuts, therefore to make it well-defined, we need to properly specify the branch. This is archived by  demanding that  the resulting  entanglement entropy  is consistent with causality and  positivity of $\Delta S[\bar{C}]$. In imposing these conditions, it is convenient to adopt the quasi-particle picture \cite{Calabrese:2007mtj,Nozaki:2014uaa,Nozaki:2014hna} for the time evolution of the entanglement entropy in a local quench. It claims the following: By a local quench,   a pair of entangled quasi-particles  is created,  one of which  is  propagating along  one spatial direction at the speed of light, and the other  along propagates along  the opposite direction. 

The change of the CFT entanglement entropy  $\Delta S[\bar{C}]$ can be non-zero only when one of such particles is in the $\bar{C}$ while the other is not \cite{Calabrese:2007mtj,Caputa:2014vaa,Nozaki:2013wia,Asplund:2014coa,Caputa:2015waa, Ugajin:2013xxa,Asplund:2013zba,Nozaki:2014hna,Nozaki:2014uaa,Caputa:2014eta,David:2016pzn,Hartman:2015lfa}.  This condition constrains possible  branches, since
vanishing of  the entanglement entropy  $\Delta S[\bar{C}] =0$  is equivalent to choose the branch where  $(z, z_{5}) \rightarrow 1$ in the $\ve \rightarrow 0$ limit . If we have multiple branches satisfying the condition, the intuition coming from the dual  holographic setup suggests that  we  should take the one giving the minimal value of $\Delta S[\bar{C}]$.

In our setup, the causality condition tells us that the CFT entanglement entropy is vanishing  in the following three cases: (1) We insert the operator in the domain of dependence of the intervals, ie $x_{0} \in D[\bar{C}_{1}]$ or $x_{0} \in D[\bar{C}_{2}]$  (2) We do not insert the operator in these domains of dependence, but the right moving particle created by the quench  enters  $D[\bar{C}_{2}]$  and the  left-mover enters $D[\bar{C}_{1}]$.  (3) We insert the operator in $D[C]$,  where $C$ is the complement of $\bar{C}$.

\begin{figure}[t]
	\centering
	\includegraphics[width=10cm]{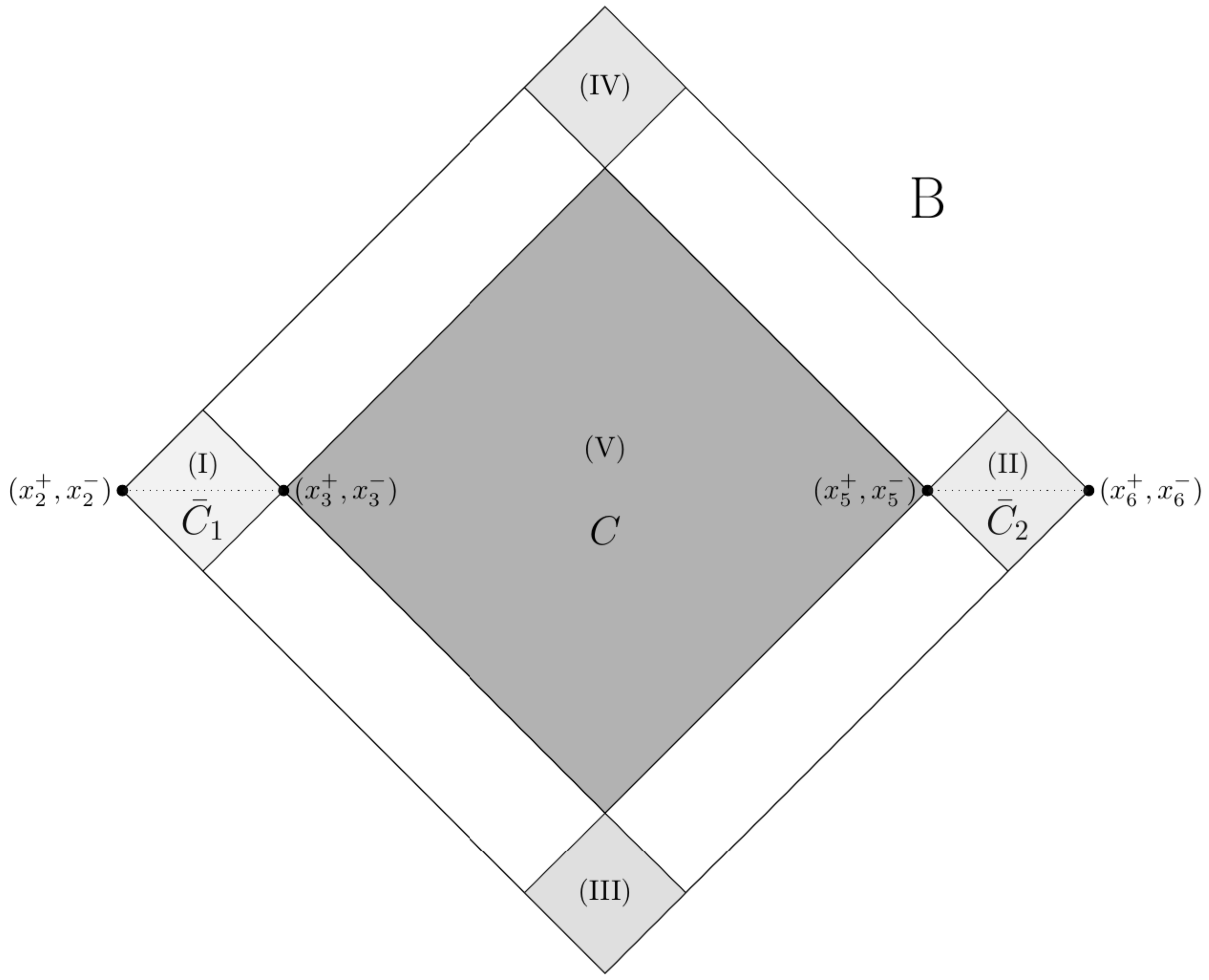}
	\hspace{0.5cm}
	
	\hspace{0.5cm}
	\caption{\small{The regions in which $\Delta S[\bar{C}] $ must vanish. The shaded regions correspond to $(I)-(V)$.}}
	\label{fig:vanishingRegion}
\end{figure}

More explicitly, the conditions that the CFT entanglement entropy must vanish in the above regions  imply we should choose  the branch$ (z^+,z^-) \to (1,1) $ and $ (\tilde{z}_5^+,\tilde{z}_5^-) \to (1,1) $ when the operator $\mathcal{O}$ is  inserted at 

\begin{itemize}
	\item[(I)]   $D[\bar{C}_{1}]$: $ x_0^+ < x_3^+$, $ x_0^- < x_3^-$,
	\item[(II)] $D[\bar{C}_{2}] $: $ x_5^+ < x_0^+$, $ x_5^- < x_0^-$,
	\item[(III)] Union of the causal pasts of $D[\bar{C}_{1}]$ and  $D[\bar{C}_{2}]$: $ x_0^+ < x_3^+$, $ x_5^- < x_0^-$,
	\item[(IV)] Union of the causal futures of $D[\bar{C}_{1}]$ and  $D[\bar{C}_{2}]$: $ x_5^+ < x_0^+$, $ x_0^- < x_3^-$,
	\item[(V)]  $D[C]$: $ x_3^+ < x_0^+<x_5^+$, $ x_3^- < x_0^-<x_5^-$.
\end{itemize}
(See figure \ref{fig:vanishingRegion}.)

The expression of $\Delta S[\bar{C}]$, when the operator is inserted in other regions, is given by suitable analytic continuations of \eqref{eq:ordinary entanglement entropy + shock part} in $x_{0}$ from the above regions $(I)-(V)$.
By using the above standard choices, we can determine branch cuts on the other regions from the consistency of analytic continuation for $ x_0^{\pm} $.

We take the region $x_{0}^{+}<x_{3}^{+}$ and $x_{3}^{-}<x_{0}^{-}<x_{5}$ as an example of such a calculation and determine possible branch choices in the region. 
Starting from the three regions (I), (III) and (V) which are adjacent to the region $x_{0}^{+}<x_{3}^{+}$ and $x_{3}^{-}<x_{0}^{-}<x_{5}^{-}$, we move the operator $ \mathcal{O} $ to the region $x_{0}^{+}<x_{3}^{+}$ and $x_{3}^{-}<x_{0}^{-}<x_{5}^{-}$. We expand $z^{\pm}$ and $\tilde{z}^{\pm}_5$ to the first order in $\ve$, which is very small compared to $\beta$, and focus on the change of their imaginary parts under the move.
For the case (I), the imaginary part of $z^-$ changes sign from plus to minus at $ x^{-}_{0} = x^{-}_{3}$, but the others not.
In such case, we choose the branches as $ (z^+,z^-) \to (1,e^{2\pi i}) $ and $ (\tilde{z}_5^+,\tilde{z}_5^-) \to (1,1) $.
For the case (III), the imaginary part of $ \tilde{z}^{-}_{5} $ changes sign from plus to minus  $ x^{-}_{0} = x^{-}_{5}$, but the others not.
Similarly, we choose the branches as $ (z^+,z^-) \to (1,1) $ and $ (\tilde{z}_5^+,\tilde{z}_5^-) \to (1,e^{2\pi i}) $.
For the case (V) , the imaginary part of $ z^{+} $ changes sign from minus to plus at  $ x^{+}_{0} = x^{+}_{3}$, but the others not. 
In this case,  we choose the branches as $ (z^+,z^-) \to (e^{2\pi i} ,1) $ and $ (\tilde{z}_5^+,\tilde{z}_5^-) \to (1,1) $.
By the above calculation, we have finished determining possible branch choices in the region $x_{0}^{+}<x_{3}^{+}$ and $x_{3}^{-}<x_{0}^{-}<x_{5}^{-}$.

Having specified the branch cuts in the region, we can calculate the analytic expression for $ \Delta S[\bar{C}] $ in the region. 
Since each branch cut gives a different $ \Delta S[\bar{C}] $ and as noted before we must pick up the dominant contribution corresponding to the minimum $\Delta S[\bar{C}]$ in the region \cite{Asplund:2014coa}.
For example, we focus on the region $x_{0}^{+}<x_{3}^{+}$ and $x_{3}^{-}<x_{0}^{-}<x_{5}$, and calculate $ \Delta S[\bar{C}] $.
In this region, $ x_0^+ < x_3^+$ and $ x_3^- < x_0^-<x_5^-$, we must compare the above three branch choices obtained from the three regions, (I), (III) and (V), and choose the minimum one.
This gives
\begin{equation}
	\begin{aligned}
		\Delta S[\bar{C}]&= \frac{c}{6} \log \left[\frac{\beta}{\pi \varepsilon} \frac{\sin \pi \alpha}{\alpha} \frac{\sinh\left(  \frac{\pi}{\beta}\left(x_{3}^{+}-x_{0}^{+}\right) \right)  \sinh \left(  \frac{\pi}{\beta}\left(x_{0}^{+}-x_{2}^{+}\right) \right) }{\sinh \left(  \frac{\pi}{\beta}\left(x_{3}^{+}-x_{2}^{+}\right)\right) }\right]\\
		&\hspace{6cm} \text{for  the region }:  x_0^+ < x_3^+ \text{ and }  x_3^- < x_0^-<x_5^-.
	\end{aligned}
\end{equation}
Similar results hold for the other regions.
By combining the above results and $ S_\beta[\bar{C}] $, we get the final expression for the entire region in the universe $B$.

Until now, we have focused on the CFT entanglement entropy of the two disjoint intervals. 
However, we easily can extend the above analysis to the single interval case by removing  either region $\bar{C}_1$ or $\bar{C}_2$ and following the similar procedure.

\bibliographystyle{JHEP}
\bibliography{Island}

\end{document}